\documentclass[letterpaper,aps,prl,reprint,superscriptaddress]{revtex4-1}

\usepackage{graphicx}
\usepackage{graphics}      % standard graphics specifications
\usepackage{longtable}     % helps with long table options
\usepackage{url}           % for on-line citations
\usepackage{bm}            % special 'bold-math' package
\usepackage{amsmath}
\usepackage{titlesec}
\pdfoutput=1

\begin{document}
	\title{Nematic Fluctuations in an Orbital Selective Superconductor Fe$_{1+y}$Te$_{1-x}$Se$_{x}$}
	
	\author{Qianni Jiang}
	\affiliation{Department of Physics, University of Washington, Seattle, Washington 98195, USA}
	\author{Yue Shi}
	\affiliation{Department of Physics, University of Washington, Seattle, Washington 98195, USA}
	\author{Morten H. Christensen}
	\affiliation{School of Physics and Astronomy, University of Minnesota, Minneapolis, Minnesota 55455, USA}
	\author{Joshua Sanchez}
	\affiliation{Department of Physics, University of Washington, Seattle, Washington 98195, USA}
	\author{Bevin Huang}
	\affiliation{Department of Physics, University of Washington, Seattle, Washington 98195, USA}
	\author{Zhong Lin}
	\affiliation{Department of Physics, University of Washington, Seattle, Washington 98195, USA}
	\author{Zhaoyu Liu}
	\affiliation{Department of Physics, University of Washington, Seattle, Washington 98195, USA}
	\author{Paul Malinowski}
	\affiliation{Department of Physics, University of Washington, Seattle, Washington 98195, USA}
	\author{Xiaodong Xu}
	\affiliation{Department of Physics, University of Washington, Seattle, Washington 98195, USA}
	\affiliation{Department of Material Science and Engineering, University of Washington, Seattle, Washington 98195, USA}
	\author{Rafael M. Fernandes}
	\affiliation{School of Physics and Astronomy, University of Minnesota, Minneapolis, Minnesota 55455, USA}
	\author{Jiun-Haw Chu}
	\affiliation{Department of Physics, University of Washington, Seattle, Washington 98195, USA}
	\date{\today}
	\begin{abstract}
		We present a systematic study of the nematic fluctuations in the iron chalcogenide superconductor Fe$_{1+y}$Te$_{1-x}$Se$_{x}$ ($0 \leq x \leq 0.53$) using the elastoresistivity technique. Near $x = 0$, in proximity to the double-stripe magnetic order of Fe$_{1+y}$Te, a diverging $B_{1g}$ nematic susceptibility is observed. Upon increasing $x$, despite the absence of magnetic order, the $B_{2g}$ nematic susceptibility increases and becomes dominant, closely following the strength of the $(\pi, \pi)$ spin fluctuations. Over a wide range of compositions ($0.17 \leq x \leq 0.53$), while the $B_{2g}$ nematic susceptibility follows a Curie temperature dependence (with zero Weiss temperature) at low temperatures, it shows deviations from Curie-Weiss behavior for temperatures higher than $50$K. This is the opposite of what is observed in typical iron pnictides, where Curie-Weiss deviations are seen at low temperatures. We attribute this unusual temperature dependence to a loss of coherence of the $d_{xy}$ orbital, which is supported by our theoretical calculations. Our results highlight the importance of orbital differentiation on the nematic properties of iron-based materials. 
	\end{abstract} 
	\maketitle
	\titlespacing\section{0pt}
\begin{figure}
		\includegraphics[trim={0 0.1cm 0cm 0.2cm},clip,width=0.5\textwidth]{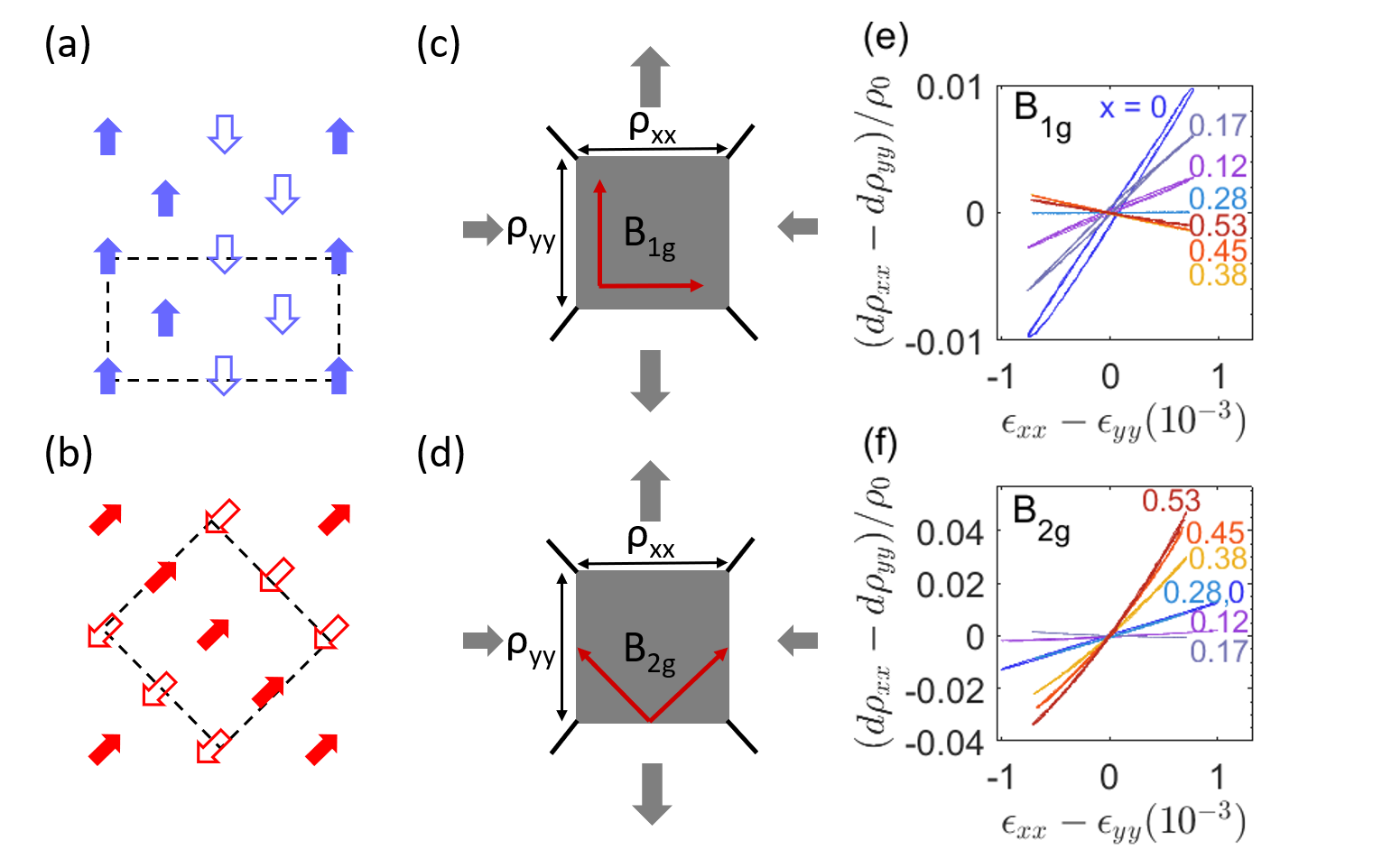}
		\caption{(a-b) Schematic spin configurations of the (a) double-stripe phase, with wave-vector $\mathbf{Q} = (\pi, 0)$, and (b) single-stripe phase, with $\mathbf{Q} = (\pi, \pi)$. (c-d) Schematic diagrams of the Montgomery method for the elastoresistivity measurement in (c) $B_{1g}$ and (d) $B_{2g}$ configuration. (e-f) The anisotropic resistivity $(\rho_{xx} - \rho_{yy})$ as a function of anisotropic strain $(\epsilon_{xx} - \epsilon_{yy})$ for Fe$_{1+y}$Te$_{1-x}$Se$_{x}$ ($0 \leq x \leq 0.53$) at $T = 20$K in (e) $B_{1g}$ and (f) $B_{2g}$ channels. The $B_{1g}$ elastoresistivity coefficient $m_{11} - m_{12}$ and $B_{2g}$ elastoresistivity coefficient $2m_{66}$ can be obtained by fitting the linear slope of resistivity versus strain. The samples with high doping concentrations ($x = 0.38, 0.45, 0.53$) show predominantly a $B_{2g}$ response while the low doping ones ($x = 0, 0.12$) show comparable $B_{1g}$ and $B_{2g}$ responses.}
		\label{fig:Fig1}
	\end{figure}

{18pt plus 4pt minus 2pt}{18pt plus 2pt minus 2pt}
	The intricate interplay between magnetism and nematicity in different families of iron-based superconductors has attracted great interest in the past few years \cite{Fernandes2014,PDai2015,Si2016}. In iron pnictides, magnetism and nematicity are tightly coupled; the antiferromagnetic transition is always coincidental with, or closely preceded by, a tetragonal-to-orthorhombic structural transition. The proximity of the two transitions can be naturally explained within the spin-nematic scenario, where the structural transition is driven by a vestigial nematic order arising from fluctuations associated with the antiferromagnetic stripe transition (see Fig. \ref{fig:Fig1}(b))\cite{CFang2008,Xu08,Fernandes2012}. In iron chalcogenides, the coupling between magnetism and nematicity is less obvious. FeSe undergoes a nematic phase transition without any long-range magnetic order \cite{McQueen2009,Bohmer2017}, which has been interpreted as evidence that the nematic order in FeSe is of orbital origin \cite{Baek2015}. Nevertheless, spin stripe fluctuations do develop below the nematic transition \cite{Wang2016}, and static stripe order can be induced by hydrostatic pressure \cite{Kothapalli16,Matsuura2017}.
	\begin{figure*}
	\includegraphics[width=1\textwidth]{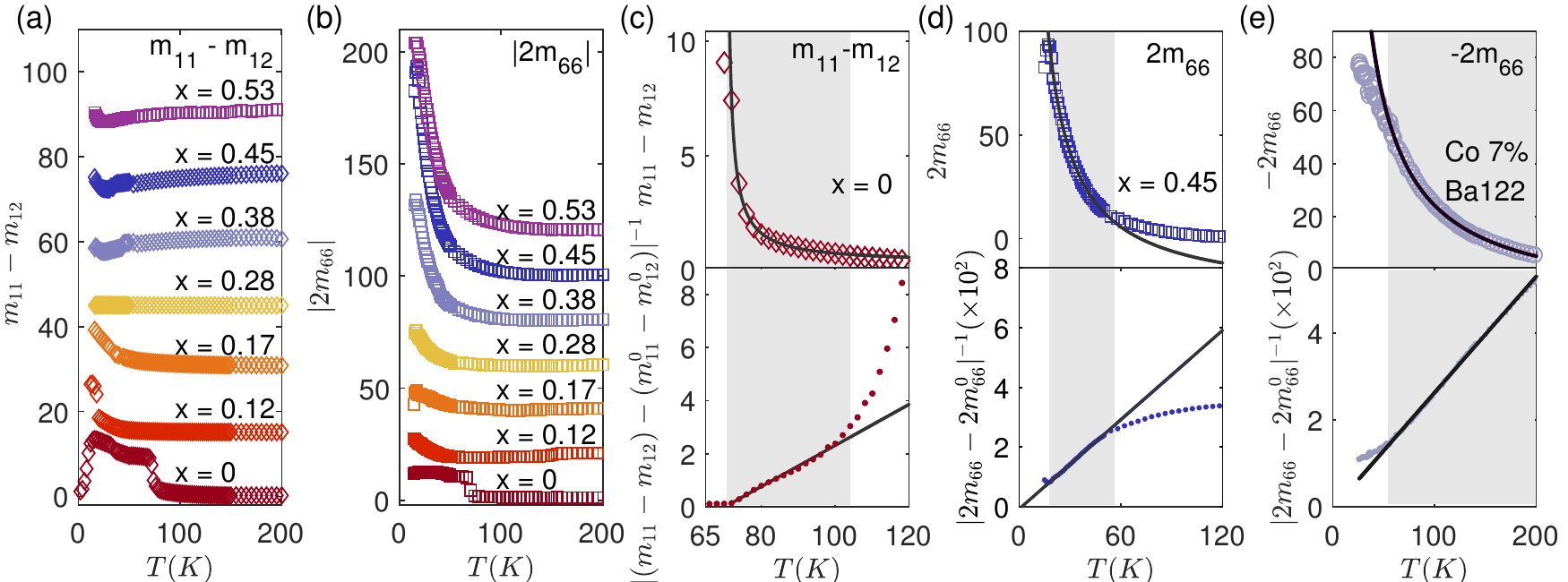}
	\caption{Temperature and doping dependence of nematic fluctuations of annealed Fe$_{1+y}$Te$_{1-x}$Se$_{x}$ in the (a) $B_{1g}$ channel, with elastoresistivity coefficient $m_{11} - m_{12}$, and (b) $B_{2g}$ channel, with  $2m_{66}$. For clarity, the elastoresistivity data for each doping are offset by 15 and 20 in (a) and (b), respectively. (c-e) Temperature dependence of (c) $m_{11} - m_{12}$ for $x = 0$, (d) $2m_{66}$ for $x = 0.45$ and (e) $-2m_{66}$ for Ba(Fe$_{0.93}$Co$_{0.07}$)$_{2}$As$_{2}$. Lower panels show the inverse. Solid black curves are Curie-Weiss fits. The optimal fitting range is determined by the greatest corresponding adjusted R-square value. Shaded gray regions indicate the range of temperatures where the elastoresistivity coefficients follow a Curie-Weiss law. }
	\label{fig:fig2}
	\end{figure*}
	\\
	\indent While there are ongoing debates on the mechanism by which nematicity forms without static magnetism in FeSe \cite{Glasbrenner2015,Wang2015,Yu2015,Khodas16,Kontani17}, Fe$_{1+y}$Te$_{1-x}$Se$_{x}$ provides another platform to approach this problem. As selenium is replaced by tellurium (i.e. $x$ is changed from $1$ to $0$), the nematic phase transition is suppressed \cite{Terao2019}, and inelastic neutron scattering experiments revealed a complex evolution of the spin correlations associated with different magnetic patterns \cite{Lumsden2010,Liu2010,Zaliznyak10316,Xu2018}. In particular, close to optimal doping ($x \sim 0.5$), the wave-vector of spin fluctuations at low temperatures is $(\pi, \pi)$ [in the crystallographic Brillouin zone], identical to the antiferromagnetic order in the iron pnictides. As the tellurium concentration increases, both superconductivity and the $(\pi, \pi)$ spin fluctuations disappear. The latter are replaced by short-range magnetic correlations near $(0, \pi)$ that eventually condense into the static double-stripe phase in Fe$_{1+y}$Te \cite{Li2009} (Fig. \ref{fig:Fig1}(a)). Previous elastoresistivity measurements revealed a diverging $B_{2g}$ nematic susceptibility in optimally doped Fe$_{1+y}$Te$_{1-x}$Se$_{x}$, consistent with the existence of $(\pi, \pi)$ spin fluctuations \cite{Kuo958}. This finding suggests that nematic and magnetic fluctuations remain strongly intertwined even in the absence of static nematic and magnetic orders. Nevertheless, in contrast to the magnetic sector, the behavior of nematic fluctuations for doping concentrations beyond optimal is still poorly characterized. The compositional dependence of the nematic susceptibility in Fe$_{1+y}$Te$_{1-x}$Se$_{x}$ would therefore constitute an important step in the effort to elucidate the relationship between nematicity and magnetism.
	\\
	\indent Another motivation to study Fe$_{1+y}$Te$_{1-x}$Se$_{x}$ is to understand the influence of orbital selectivity on the nematic instability. Orbital selectivity (or orbital differentiation) refers to the fact that different orbitals are renormalized differently by electronic correlations, a characteristic property of Hund's metals that appears to be much more prominent in the iron chalcogenides in comparison with the pnictides \cite{Yin2011,Lanata2013,Medici2014}. Experimentally, recent scanning tunneling microscopy (STM) measurements revealed the impact of orbital differentiation on the superconducting state \cite{Sprau75}. Theoretically, it has been suggested that orbitally selective spin fluctuations may be the origin of nematicity without magnetism in FeSe \cite{Fanfarillo2018}. Nematic order was also proposed to enhance orbital selectivity by breaking the orbital degeneracy, leading to asymmetric effective masses in different $d$-orbitals \cite{Yu2018}. The effect of orbital differentiation becomes even more extreme as selenium is replaced by tellurium. In Fe$_{1+y}$Te$_{1-x}$Se$_{x}$, angle resolved photoemission spectroscopy (ARPES) revealed a strong loss of spectral weight of the $d_{xy}$ orbital at high temperatures, which was interpreted in terms of proximity to an orbital-selective Mott transition \cite{Yi2015}. Similar drastic changes were also observed as a function of doping \cite{Liu2015}, mimicking the evolution of spin fluctuations. Nevertheless, the impact of orbital incoherence on nematicity remains little explored \cite{Bascones17}.
	\\
	\indent In this report, we present systematic measurements of both the $B_{1g}$ and $B_{2g}$ nematic susceptibilities of Fe$_{1+y}$Te$_{1-x}$Se$_{x}$ ($0 \leq x \leq 0.53$) using the elastoresistivity technique. We demonstrate that the doping dependence of the two nematic susceptibilities closely track the evolution of the corresponding spin fluctuations. In particular, a diverging $B_{1g}$ nematic susceptibility is observed in the parent compound Fe$_{1+y}$Te, suggesting that the spin-nematic paradigm also applies to the double-stripe AFM order \cite{Zhang2017, Moreo17, Borisov2019}. A diverging $B_{2g}$ nematic susceptibility is observed over a wide range of doping ($0.17 \leq x \leq 0.53$), and its magnitude is strongly enhanced by both Se doping and annealing. In addition, the temperature dependence of the $B_{2g}$ nematic susceptibility shows significant deviation from Curie-Weiss behavior above 50K. This is in sharp contrast to the iron pnictides, where the Curie-Weiss temperature dependence extends all the way to 200K. This unusual temperature dependence is captured by a theoretical calculation that includes the loss of spectral weight of the $d_{xy}$ orbital, revealing its importance for $B_{2g}$ nematic instability.
	\\
	\indent Single crystals of Fe$_{1+y}$Te$_{1-x}$Se$_{x}$ were grown by the modified Bridgeman method. The electrical, magnetic and superconducting properties of Fe$_{1+y}$Te$_{1-x}$Se$_{x}$ are known to sensitively depend on $y$, the amount of excess iron. To study these effects, crystals were annealed in selenium vapor to reduce the amount of excess iron. By symmetry, the $B_{1g}$ and $B_{2g}$ nematic susceptibilities are proportional to the elastoresistivity coefficients $m_{11} - m_{12}$ and $2m_{66}$, respectively. We performed the elastoresistivity measurements in the Montgomery geometry, which enables simultaneous determination of the full resistivity tensor, hence the precise decomposition into different symmetry channels, as illustrated in Fig. \ref{fig:Fig1}(c) and (d). Details of the Montgomery elastoresistivity measurements can be found elsewhere \cite{Kuo958}. The crystal orientation is determined by polarization resolved Raman spectroscopy. Representative data of anisotropic resistivity as a function of anisotropic strain at 20K in $B_{1g}$ and $B_{2g}$ channels are shown in Fig. \ref{fig:Fig1}(e) and (f). The $B_{1g}$ elastoresistivity coefficient $m_{11} - m_{12}$ and the $B_{2g}$ one,  $2m_{66}$, can be obtained by fitting the linear slope of resistivity versus strain. Samples with high doping concentrations ($x = 0.38, 0.45, 0.53$) show predominantly a $B_{2g}$ response while the low doping ones ($x = 0, 0.12$) show comparable $B_{1g}$ and $B_{2g}$ responses.
	\\
	\indent Figs. \ref{fig:fig2}(a) and (b) show the temperature dependence of $m_{11} - m_{12}$ and  $2m_{66}$ of annealed Fe$_{1+y}$Te$_{1-x}$Se$_{x}$ for $0 \leq x \leq 0.53$. For $0.28 \leq x \leq 0.53$, $2m_{66}$ shows a strong temperature dependence that grows continuously as temperature decreases. For $x = 0.45$, $2m_{66}$ reaches a value of $\sim$ 100, comparable to optimally doped pnictides. While preserving a similar diverging temperature dependence, the maximum absolute value of $2m_{66}$ decreases rapidly as selenium concentration decreases, from 100 for $x = 0.45$ to 8 for $x = 0.17$. On the other hand, $m_{11} - m_{12}$ shows a diverging response when $x$ is below $0.17$, which is in the vicinity of the double-stripe AFM order. As selenium concentration increases, $m_{11} - m_{12}$ evolves to a temperature independent response, with small kinks at low temperatures likely coming from contamination of $2m_{66}$ due to misalignment. Overall, our observation of the doping dependence of $2m_{66}$ and $m_{11} - m_{12}$ is consistent with the evolution of low-temperature spin fluctuations from predominantly $(\pi,0)$ at small $x$ to predominantly $(\pi,\pi)$ at optimal doping $x\sim 0.5$ \cite{Lumsden2010,Liu2010,Zaliznyak10316,Xu2018}.
	\\
	\indent To gain more insight, we fit the $2m_{66}$ and $m_{11} - m_{12}$ to a Curie-Weiss temperature dependence:
	\begin{eqnarray}
	m=m^{0}+\frac{\lambda}{a(T-T^{*})}
	\end{eqnarray}
	For the parent compound Fe$_{1+y}$Te, $m_{11} - m_{12}$ can be well fitted to a Curie-Weiss behavior in the temperature range just above the double stripe AFM ordering temperature $T_{\rm mag}$ = 71.5K (Fig. \ref{fig:fig2}(c)). The fitted Curie-Weiss temperature $T^{*}$ is slightly smaller than $T_{\rm mag}$. Despite the smaller absolute value ($\sim$ 10 at maximum), the behavior of $m_{11} - m_{12}$ is reminiscent of the $2m_{66}$ in the parent phase of iron pnictides, suggesting that the spin-nematic mechanism is still at play here, in agreement with theoretical expectations \cite{Zhang2017, Moreo17, Borisov2019}. 
	\\
	\begin{figure}
		\includegraphics[trim={0 1cm 0 0cm},clip,width=0.48\textwidth]{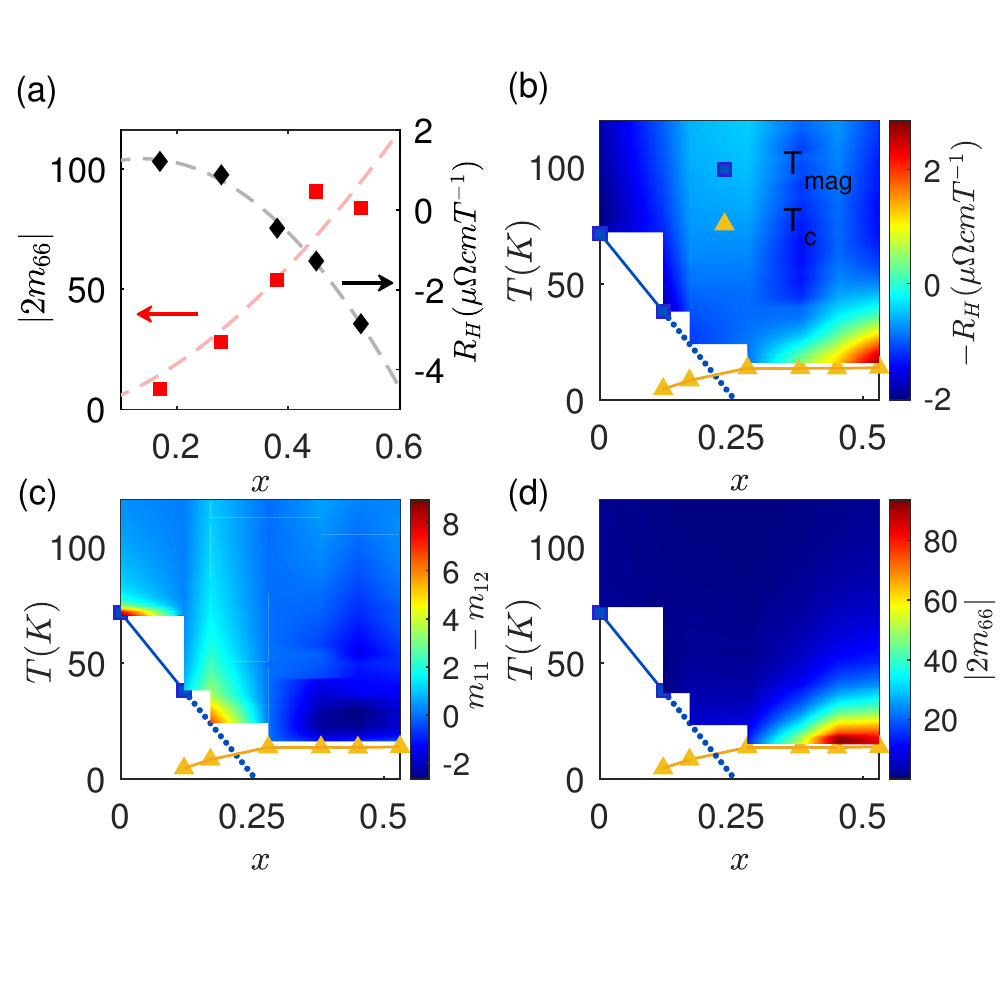}
		\caption{Comparison with the Hall coefficient $R_{H}$. (a) Doping dependence of the absolute value of the $B_{2g}$ elastoresistivity coefficient $2m_{66}$ (red squares) and of $R_{H}$ (black diamonds) at 16K. Dashed lines are guide to the eyes. (b) Colormap of the negative Hall coefficient -$R_{H}$, (c) of $m_{11} - m_{12}$ (d) and of $|2m_{66}|$ as a function of temperature and doping. The double-spin stripe and the superconducting transition temperatures are denoted as blue squares and yellow triangles, respectively.}
		\label{fig:fig3}
	\end{figure}
	\indent Fig. \ref{fig:fig2}(d) shows the Curie-Weiss fitting of $2m_{66}$ for the $x = 0.45$ sample. The fitting of $2m_{66}$ only works at low temperatures, as can be seen in the linear temperature dependence of $|2m_{66} -2m_{66}^{0}|^{-1}$ below 50K. It shows a significant deviation for temperature greater than 50K. The $T^{*}$ obtained from the low-temperature fitting is close to 0K.  Intriguingly, the $T^{*}$ extracted from the Curie-Weiss fitting is approximately zero for all $0.17 \leq x \leq 0.53$, while the Curie constant $\lambda/a$ decreases with $x$ (SOM). While the number of doping concentrations studied in the current work is insufficient to support a power law analysis, $2m_{66}$ at constant $T = 16$K appears to be diverging as $x$ increases from 0.17 to 0.45 (Fig. \ref{fig:fig3}(a)). Both the doping dependence and the near zero $T^{*}$ are consistent with the existence of a putative nematic quantum critical point at $x \sim 0.5$. Interestingly, recent work doping FeSe with Te suggests that the $90$K nematic transition of FeSe is continuously suppressed and extrapolates to $0$ at $x \sim 0.5$ \cite{Terao2019}.
%Recent successes in the synthesis of single crystals and thin films bridge the phase diagram of Fe$_{1+y}$Te$_{1-x}$Se$_{x}$ from $x = \sim 0.5$ to $x = 1$, the end member FeSe. The 90K nematic transition in FeSe is continuously suppressed as $x$ decreases from 1, and a nematic quantum critical point can be extrapolated to $x \sim 0.5$. 	
	\\
	\indent This deviation from Curie-Weiss at high temperatures is very unusual. In the iron pnicitides, such a deviation was only observed at low temperatures in transition-metal doped BaFe$_{2}$As$_{2}$ (Fig. \ref{fig:fig2}(e)) and LaFeAsO. This unusual temperature dependence of $2m_{66}$ appears to echo the coherent-incoherent crossover observed by ARPES \cite{Yi2015}, where the spectral weight of the $d_{xy}$ orbital is strongly suppressed as the temperature increases or as the selenium concentration decreases. To further confirm this correlation, we measured the Hall coefficient $R_H$, which has been demonstrated to be a good indicator of this incoherent-to-coherent crossover \cite{Ding2014,Otsuka2019,Liu2015}. The recovery of the $d_{xy}$ spectral weight is generally correlated with a sign-change of $R_H$ \cite{Otsuka2019} from positive to negative. Fig. \ref{fig:fig3}(a) shows the low-temperature $R_H$ and $2m_{66}$ as a function of doping, whereas Fig. \ref{fig:fig3}(b-d) contain the full temperature and doping dependence of $R_H$,  $m_{11} - m_{12}$, and $|2m_{66}|$, respectively. These plots reveal the strong correlation between a negative $R_H$ and an enhancement of $2m_{66}$.
	\\
		\begin{figure}
		\includegraphics[width=0.38\textwidth]{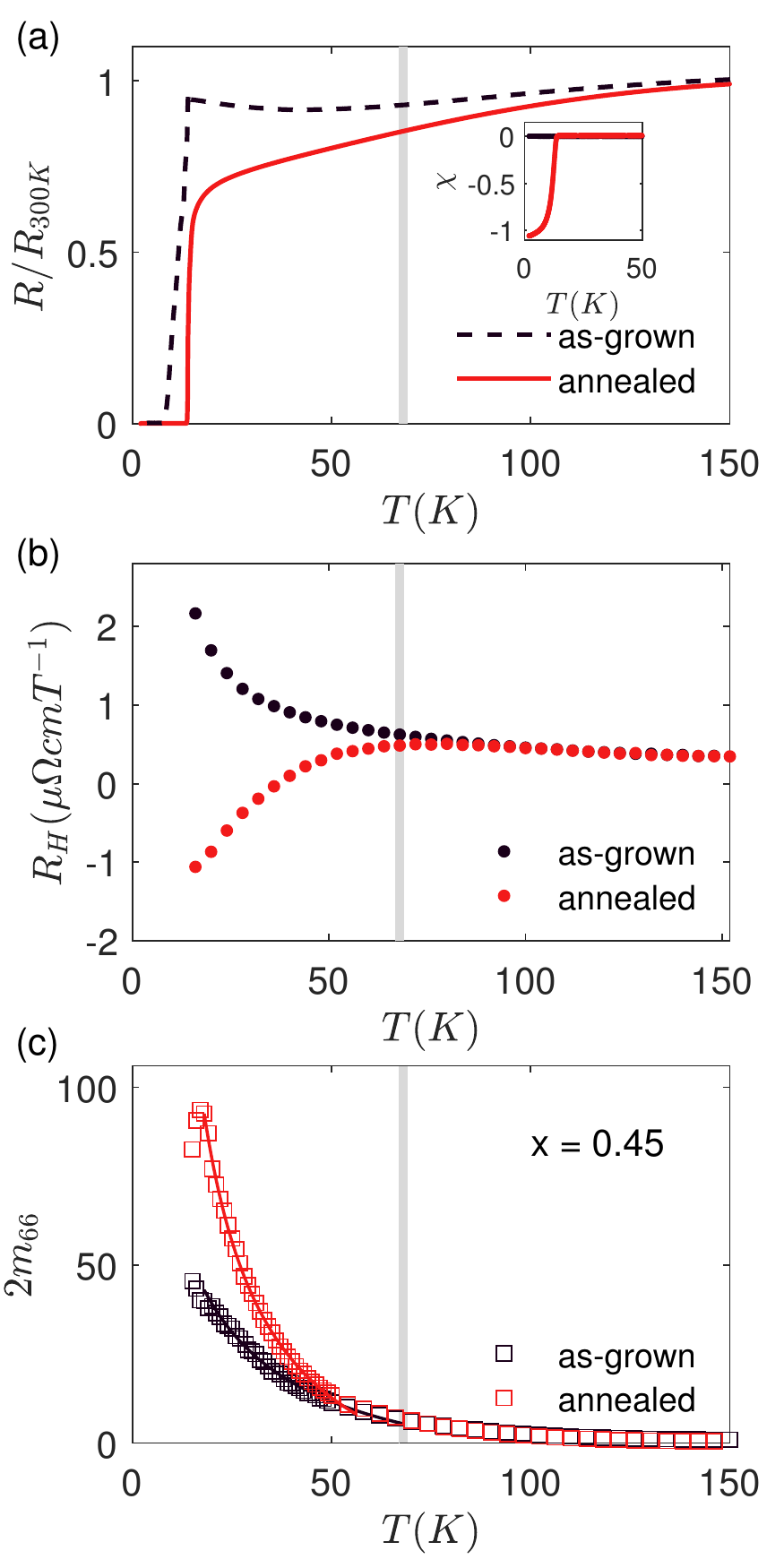}
		\caption{The effect of annealing on the nematic susceptibility of Fe$_{1+y}$Te$_{1-x}$Se$_{x}$. (a-c) Temperature dependence of (a) normalized in-plane resistivity ($R/R_{300 \rm{K}}$), (b) Hall coefficient $	R_{H}$ and (c) elastoresistivity coefficient $2m_{66}$ of as-grown (black) and annealed (red) samples for $x = 0.45$. The vertical grey line marks the temperature below which the behavior of annealed and as-grown samples starts to deviate from each other. Inset of (a) shows the temperature dependence of the zero field cooling (ZFC) magnetic susceptibility measured at 100Oe (H $\parallel$ ab). The superconducting volume fraction is significantly enhanced for the annealed sample. }
		\label{fig:Fig4}
	\end{figure}
	\indent The properties of Fe$_{1+y}$Te$_{1-x}$Se$_{x}$ also depend on the amount of excess iron, which can only be removed by annealing \cite{Sun_2019}. Taking $x = 0.45$ as an example, the resistance of the annealed sample is metallic for temperatures below 150K (Fig. \ref{fig:Fig4}(a)). As Fig. \ref{fig:Fig4}(b) shows, at around 40K the Hall coefficient of the annealed sample turns from positive to negative, which is a signature of incoherent to coherent crossover. In contrast, the resistance of the as-grown sample shows a weakly insulating upturn at low temperatures (Fig. \ref{fig:Fig4}(a) black dashed curve), and the Hall coefficient remains positive at all temperatures (Fig. \ref{fig:Fig4}(b) black circles), indicating that the $d_{xy}$ orbital is still incoherent at low temperatures. Interestingly, at the same temperature where the resistance and the Hall coefficient of the as-grown and annealed samples depart from each other, the elastoresistivity coefficient $2m_{66}$ shows a pronounced enhancement for the annealed sample (Fig. \ref{fig:Fig4}(c)). Such an enhancement was observed in all annealed samples (SOM), providing further evidence of the correlation between the enhancement of the nematic susceptibility and the coherence of the $d_{xy}$ orbital.
	\\
	\indent The doping and annealing dependences of $2m_{66}$ presented above strongly suggest that the $B_{2g}$ nematic susceptibility also have an orbitally-selective character. %This is in agreement with a recent theoretical calculation that reveals a strongly orbitally dependent nematic susceptibility. In particular, the $d_{xy}$ orbital contributes most to the overall nematic instability \cite{Christensen2016}. To gain more insight, we calculated the nematic susceptibility with and without a reduced spectral weight in the $d_{xy}$ orbital to simulate the orbital correlation effect in Fe$_{1+y}$Te$_{1-x}$Se$_{x}$ and BaFe$_{2}$(As$_{1-x}$P$_{x}$)$_{2}$, respectively. The calculated inverse of the $B_{2g}$ nematic susceptibility is plotted in FIG. \ref{fig:fig5}(d). The excellent agreement with the experimentally measured $|2m_{66} - 2m_{66}^{0}|^{-1}$ (Fig. \ref{fig:fig5}(b)) confirms the orbitally selective nature of the nematic susceptibility.  
	Indeed, previous theoretical works have highlighted the impact of orbital degrees of freedom on spin-driven nematicity \cite{Fanfarillo15, Khodas16, Kontani16, Yu2018, Chubukov18, Fanfarillo2018}. Using a slave-spin approach, Ref. \cite{Bascones17} found a suppression of the orbital-nematic susceptibility due to orbital incoherence. To model our data, we employ the generalized random phase approximation (RPA) of Ref. \cite{Christensen2016} to compute the spin-driven nematic susceptibility for the five-orbital Hubbard-Kanamori model (details in the SM). For fully coherent orbitals, it was found that the largest contribution to the nematic susceptibility $\chi_{\rm nem}$ comes from the $d_{xy}$ orbital. Thus, one expects that $\chi_{\rm nem}$ would be suppressed if the $d_{xy}$ orbital were to become less coherent.
	\\
		\begin{figure}
		\includegraphics[trim={0 0cm 0cm 0cm},clip,width=0.5\textwidth]{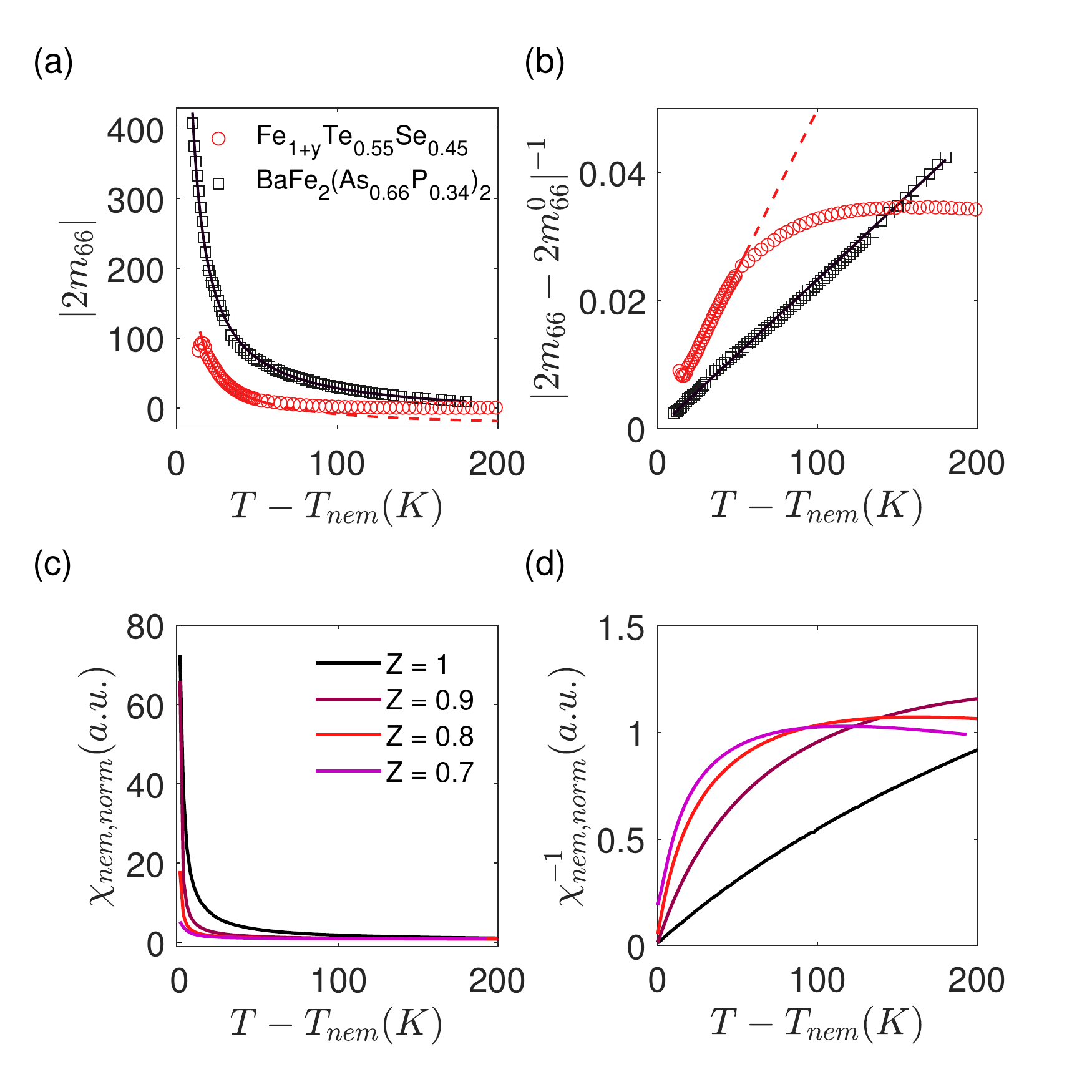}
		\caption{Comparison with theoretical calculations. (a-b) Temperature dependence of (a) $|2m_{66}|$ and (b) $|2m_{66} - 2m_{66}^{0}|^{-1}$ of optimally doped Fe$_{1+y}$Te$_{0.55}$Se$_{0.45}$ (red circles) and BaFe$_{2}$(As$_{0.66}$P$_{0.34}$)$_{2}$ (black squares). The red and black lines show Curie Weiss fittings. The data for BaFe$_{2}$(As$_{0.66}$P$_{0.34}$)$_{2}$ follows a Curie-Weiss behavior all the way up to $200$K, whereas for Fe$_{1+y}$Te$_{0.55}$Se$_{0.45}$, it deviates from Curie-Weiss behavior above $\sim$ 50K. (c-d) Theoretical calculation of the normalized nematic susceptibility $\chi_{\rm nem}$ and its inverse, plotted as a function of the relative temperature with respect to the theoretical nematic transition temperature ($T-T_{\rm nem}$) for different spectral weight $0.7 \leq Z_{xy} \leq 1$.}
		\label{fig:fig5}
	\end{figure}
	\indent To verify this scenario, we calculated how $\chi_{\rm nem}$ changes upon suppressing the spectral weight $Z_{xy}$ of the $d_{xy}$ orbital. For our purposes, the reduction in $Z_{xy}$ acts phenomenologically as a proxy of the incoherence of this orbital, similarly to \cite{Sprau75}, but its microscopic origin is not important. Fig. \ref{fig:fig5}(c)-(d) contrasts the nematic susceptibility for $ 0.7 \leq Z_{xy} \leq 1$. We note two main trends arising from the suppression of $d_{xy}$ spectral weight: first, as anticipated, the nematic susceptibility (and the underlying nematic transition temperature, which is non-zero in the model) is reduced (Fig. \ref{fig:fig5}(c)). Second, its temperature dependence changes from a Curie-Weiss-like behavior over an extended temperature range to a behavior in which the inverse nematic susceptibility quickly saturates and strongly deviates from a linear-in-T dependence already quite close to the nematic transition (Fig. \ref{fig:fig5}(d)). These behaviors are remarkably similar to those displayed by the elastoresistance data shown in Fig. \ref{fig:fig5}(a)-(b), with $Z_{xy} = 1$ mimicking the behavior of optimally P-doped BaFe$_{2}$As$_{2}$ and $Z_{xy} < 1$, of optimally doped Fe$_{1+y}$Te$_{1-x}$Se$_x$. Interestingly, the susceptibility associated with $(\pi, \pi)$ fluctuations is also suppressed by the decrease in $Z_{xy}$, in qualitative agreement with the neutron scattering experiments \cite{Xu2012} (for a more detailed discussion, see SM). Of course, since $Z_{xy}$ in our model is an input, and not calculated microscopically, our model is useful to capture tendencies, but not to extract the experimental value of $Z_{xy}$. Furthermore, note that in our calculation $Z_{xy}$ is temperature-independent, while in the experiment it changes with temperature.
	\\
	\indent In summary, our results reveal the close connection between nematic fluctuations and spin fluctuations in Fe$_{1+y}$Te$_{1-x}$Se$_{x}$ for both $B_{1g}$ and $B_{2g}$ channels. Additionally, the unusual temperature dependence of the $B_{2g}$ nematic susceptibility can be attributed to the coherent-to-incoherent crossover experienced by the $d_{xy}$ orbital, providing direct evidence for the orbital selectivity of the nematic instability. Our work presents Fe$_{1+y}$Te$_{1-x}$Se$_{x}$ as an ideal platform to study the physics of intertwined orders in a strongly correlated Hund's metal.

\begin{acknowledgments}
We thank Ming Yi for fruitful discussion. The work at UW was supported by NSF MRSEC at UW (DMR-1719797). The material synthesis was supported by the Northwest Institute for Materials Physics, Chemistry, and Technology (NW IMPACT) and the Gordon and Betty Moore Foundation’s EPiQS Initiative, Grant GBMF6759 to J.-H.C.  J.H.C. acknowledge the support of the David and Lucile Packard Foundation, the Alford P. Sloan Foundation and the State of Washington funded Clean Energy Institute. Theory work (M.H.C. and R.M.F.) was supported by the U. S. Department of Energy, Office of Science, Basic Energy Sciences, Materials Sciences and Engineering Division, under Award No. DE-SC0020045. The Raman measurement is partially supported by Department of Energy, Basic Energy Sciences, Materials Sciences and Engineering Division (DE-SC0012509).
\end{acknowledgments}

\bibliography{ms}

\begin{thebibliography}{44}%
\makeatletter
\providecommand \@ifxundefined [1]{%
 \@ifx{#1\undefined}
}%
\providecommand \@ifnum [1]{%
 \ifnum #1\expandafter \@firstoftwo
 \else \expandafter \@secondoftwo
 \fi
}%
\providecommand \@ifx [1]{%
 \ifx #1\expandafter \@firstoftwo
 \else \expandafter \@secondoftwo
 \fi
}%
\providecommand \natexlab [1]{#1}%
\providecommand \enquote  [1]{``#1''}%
\providecommand \bibnamefont  [1]{#1}%
\providecommand \bibfnamefont [1]{#1}%
\providecommand \citenamefont [1]{#1}%
\providecommand \href@noop [0]{\@secondoftwo}%
\providecommand \href [0]{\begingroup \@sanitize@url \@href}%
\providecommand \@href[1]{\@@startlink{#1}\@@href}%
\providecommand \@@href[1]{\endgroup#1\@@endlink}%
\providecommand \@sanitize@url [0]{\catcode `\\12\catcode `\$12\catcode
  `\&12\catcode `\#12\catcode `\^12\catcode `\_12\catcode `\%12\relax}%
\providecommand \@@startlink[1]{}%
\providecommand \@@endlink[0]{}%
\providecommand \url  [0]{\begingroup\@sanitize@url \@url }%
\providecommand \@url [1]{\endgroup\@href {#1}{\urlprefix }}%
\providecommand \urlprefix  [0]{URL }%
\providecommand \Eprint [0]{\href }%
\providecommand \doibase [0]{http://dx.doi.org/}%
\providecommand \selectlanguage [0]{\@gobble}%
\providecommand \bibinfo  [0]{\@secondoftwo}%
\providecommand \bibfield  [0]{\@secondoftwo}%
\providecommand \translation [1]{[#1]}%
\providecommand \BibitemOpen [0]{}%
\providecommand \bibitemStop [0]{}%
\providecommand \bibitemNoStop [0]{.\EOS\space}%
\providecommand \EOS [0]{\spacefactor3000\relax}%
\providecommand \BibitemShut  [1]{\csname bibitem#1\endcsname}%
\let\auto@bib@innerbib\@empty
%</preamble>
\bibitem [{\citenamefont {Fernandes}\ \emph {et~al.}(2014)\citenamefont
  {Fernandes}, \citenamefont {Chubukov},\ and\ \citenamefont
  {Schmalian}}]{Fernandes2014}%
  \BibitemOpen
  \bibfield  {author} {\bibinfo {author} {\bibfnamefont {R.~M.}\ \bibnamefont
  {Fernandes}}, \bibinfo {author} {\bibfnamefont {A.~V.}\ \bibnamefont
  {Chubukov}}, \ and\ \bibinfo {author} {\bibfnamefont {J.}~\bibnamefont
  {Schmalian}},\ }\href {\doibase 10.1038/nphys2877} {\bibfield  {journal}
  {\bibinfo  {journal} {Nature Physics}\ }\textbf {\bibinfo {volume} {10}},\
  \bibinfo {pages} {97} (\bibinfo {year} {2014})}\BibitemShut {NoStop}%
\bibitem [{\citenamefont {Dai}(2015)}]{PDai2015}%
  \BibitemOpen
  \bibfield  {author} {\bibinfo {author} {\bibfnamefont {P.}~\bibnamefont
  {Dai}},\ }\href {\doibase 10.1103/RevModPhys.87.855} {\bibfield  {journal}
  {\bibinfo  {journal} {Rev. Mod. Phys.}\ }\textbf {\bibinfo {volume} {87}},\
  \bibinfo {pages} {855} (\bibinfo {year} {2015})}\BibitemShut {NoStop}%
\bibitem [{\citenamefont {Si}\ \emph {et~al.}(2016)\citenamefont {Si},
  \citenamefont {Yu},\ and\ \citenamefont {Abrahams}}]{Si2016}%
  \BibitemOpen
  \bibfield  {author} {\bibinfo {author} {\bibfnamefont {Q.}~\bibnamefont
  {Si}}, \bibinfo {author} {\bibfnamefont {R.}~\bibnamefont {Yu}}, \ and\
  \bibinfo {author} {\bibfnamefont {E.}~\bibnamefont {Abrahams}},\ }\href
  {\doibase 10.1038/natrevmats.2016.17} {\bibfield  {journal} {\bibinfo
  {journal} {Nature Reviews Materials}\ }\textbf {\bibinfo {volume} {1}},\
  \bibinfo {pages} {16017} (\bibinfo {year} {2016})}\BibitemShut {NoStop}%
\bibitem [{\citenamefont {Fang}\ \emph {et~al.}(2008)\citenamefont {Fang},
  \citenamefont {Yao}, \citenamefont {Tsai}, \citenamefont {Hu},\ and\
  \citenamefont {Kivelson}}]{CFang2008}%
  \BibitemOpen
  \bibfield  {author} {\bibinfo {author} {\bibfnamefont {C.}~\bibnamefont
  {Fang}}, \bibinfo {author} {\bibfnamefont {H.}~\bibnamefont {Yao}}, \bibinfo
  {author} {\bibfnamefont {W.-F.}\ \bibnamefont {Tsai}}, \bibinfo {author}
  {\bibfnamefont {J.}~\bibnamefont {Hu}}, \ and\ \bibinfo {author}
  {\bibfnamefont {S.~A.}\ \bibnamefont {Kivelson}},\ }\href {\doibase
  10.1103/PhysRevB.77.224509} {\bibfield  {journal} {\bibinfo  {journal} {Phys.
  Rev. B}\ }\textbf {\bibinfo {volume} {77}},\ \bibinfo {pages} {224509}
  (\bibinfo {year} {2008})}\BibitemShut {NoStop}%
\bibitem [{\citenamefont {Xu}\ \emph {et~al.}(2008)\citenamefont {Xu},
  \citenamefont {M\"uller},\ and\ \citenamefont {Sachdev}}]{Xu08}%
  \BibitemOpen
  \bibfield  {author} {\bibinfo {author} {\bibfnamefont {C.}~\bibnamefont
  {Xu}}, \bibinfo {author} {\bibfnamefont {M.}~\bibnamefont {M\"uller}}, \ and\
  \bibinfo {author} {\bibfnamefont {S.}~\bibnamefont {Sachdev}},\ }\href
  {\doibase 10.1103/PhysRevB.78.020501} {\bibfield  {journal} {\bibinfo
  {journal} {Phys. Rev. B}\ }\textbf {\bibinfo {volume} {78}},\ \bibinfo
  {pages} {020501} (\bibinfo {year} {2008})}\BibitemShut {NoStop}%
\bibitem [{\citenamefont {Fernandes}\ \emph {et~al.}(2012)\citenamefont
  {Fernandes}, \citenamefont {Chubukov}, \citenamefont {Knolle}, \citenamefont
  {Eremin},\ and\ \citenamefont {Schmalian}}]{Fernandes2012}%
  \BibitemOpen
  \bibfield  {author} {\bibinfo {author} {\bibfnamefont {R.~M.}\ \bibnamefont
  {Fernandes}}, \bibinfo {author} {\bibfnamefont {A.~V.}\ \bibnamefont
  {Chubukov}}, \bibinfo {author} {\bibfnamefont {J.}~\bibnamefont {Knolle}},
  \bibinfo {author} {\bibfnamefont {I.}~\bibnamefont {Eremin}}, \ and\ \bibinfo
  {author} {\bibfnamefont {J.}~\bibnamefont {Schmalian}},\ }\href {\doibase
  10.1103/PhysRevB.85.024534} {\bibfield  {journal} {\bibinfo  {journal} {Phys.
  Rev. B}\ }\textbf {\bibinfo {volume} {85}},\ \bibinfo {pages} {024534}
  (\bibinfo {year} {2012})}\BibitemShut {NoStop}%
\bibitem [{\citenamefont {McQueen}\ \emph {et~al.}(2009)\citenamefont
  {McQueen}, \citenamefont {Williams}, \citenamefont {Stephens}, \citenamefont
  {Tao}, \citenamefont {Zhu}, \citenamefont {Ksenofontov}, \citenamefont
  {Casper}, \citenamefont {Felser},\ and\ \citenamefont {Cava}}]{McQueen2009}%
  \BibitemOpen
  \bibfield  {author} {\bibinfo {author} {\bibfnamefont {T.~M.}\ \bibnamefont
  {McQueen}}, \bibinfo {author} {\bibfnamefont {A.~J.}\ \bibnamefont
  {Williams}}, \bibinfo {author} {\bibfnamefont {P.~W.}\ \bibnamefont
  {Stephens}}, \bibinfo {author} {\bibfnamefont {J.}~\bibnamefont {Tao}},
  \bibinfo {author} {\bibfnamefont {Y.}~\bibnamefont {Zhu}}, \bibinfo {author}
  {\bibfnamefont {V.}~\bibnamefont {Ksenofontov}}, \bibinfo {author}
  {\bibfnamefont {F.}~\bibnamefont {Casper}}, \bibinfo {author} {\bibfnamefont
  {C.}~\bibnamefont {Felser}}, \ and\ \bibinfo {author} {\bibfnamefont {R.~J.}\
  \bibnamefont {Cava}},\ }\href {\doibase 10.1103/PhysRevLett.103.057002}
  {\bibfield  {journal} {\bibinfo  {journal} {Phys. Rev. Lett.}\ }\textbf
  {\bibinfo {volume} {103}},\ \bibinfo {pages} {057002} (\bibinfo {year}
  {2009})}\BibitemShut {NoStop}%
\bibitem [{\citenamefont {B{\"o}hmer}\ and\ \citenamefont
  {Kreisel}(2017)}]{Bohmer2017}%
  \BibitemOpen
  \bibfield  {author} {\bibinfo {author} {\bibfnamefont {A.~E.}\ \bibnamefont
  {B{\"o}hmer}}\ and\ \bibinfo {author} {\bibfnamefont {A.}~\bibnamefont
  {Kreisel}},\ }\href@noop {} {\bibfield  {journal} {\bibinfo  {journal}
  {Journal of Physics:Condensed Matter}\ }\textbf {\bibinfo {volume} {30}},\
  \bibinfo {pages} {023001} (\bibinfo {year} {2017})}\BibitemShut {NoStop}%
\bibitem [{\citenamefont {Baek}\ \emph {et~al.}(2015)\citenamefont {Baek},
  \citenamefont {Efremov}, \citenamefont {Ok}, \citenamefont {Kim},
  \citenamefont {van~den Brink},\ and\ \citenamefont {B{\"u}chner}}]{Baek2015}%
  \BibitemOpen
  \bibfield  {author} {\bibinfo {author} {\bibfnamefont {S.-H.}\ \bibnamefont
  {Baek}}, \bibinfo {author} {\bibfnamefont {D.~V.}\ \bibnamefont {Efremov}},
  \bibinfo {author} {\bibfnamefont {J.~M.}\ \bibnamefont {Ok}}, \bibinfo
  {author} {\bibfnamefont {J.~S.}\ \bibnamefont {Kim}}, \bibinfo {author}
  {\bibfnamefont {J.}~\bibnamefont {van~den Brink}}, \ and\ \bibinfo {author}
  {\bibfnamefont {B.}~\bibnamefont {B{\"u}chner}},\ }\href {\doibase
  10.1038/nmat4138} {\bibfield  {journal} {\bibinfo  {journal} {Nature
  Materials}\ }\textbf {\bibinfo {volume} {14}},\ \bibinfo {pages} {210}
  (\bibinfo {year} {2015})}\BibitemShut {NoStop}%
\bibitem [{\citenamefont {Wang}\ \emph {et~al.}(2016)\citenamefont {Wang},
  \citenamefont {Shen}, \citenamefont {Pan}, \citenamefont {Hao}, \citenamefont
  {Ma}, \citenamefont {Zhou}, \citenamefont {Steffens}, \citenamefont
  {Schmalzl}, \citenamefont {Forrest}, \citenamefont {Abdel-Hafiez},
  \citenamefont {Chen}, \citenamefont {Chareev}, \citenamefont {Vasiliev},
  \citenamefont {Bourges}, \citenamefont {Sidis}, \citenamefont {Cao},\ and\
  \citenamefont {Zhao}}]{Wang2016}%
  \BibitemOpen
  \bibfield  {author} {\bibinfo {author} {\bibfnamefont {Q.}~\bibnamefont
  {Wang}}, \bibinfo {author} {\bibfnamefont {Y.}~\bibnamefont {Shen}}, \bibinfo
  {author} {\bibfnamefont {B.}~\bibnamefont {Pan}}, \bibinfo {author}
  {\bibfnamefont {Y.}~\bibnamefont {Hao}}, \bibinfo {author} {\bibfnamefont
  {M.}~\bibnamefont {Ma}}, \bibinfo {author} {\bibfnamefont {F.}~\bibnamefont
  {Zhou}}, \bibinfo {author} {\bibfnamefont {P.}~\bibnamefont {Steffens}},
  \bibinfo {author} {\bibfnamefont {K.}~\bibnamefont {Schmalzl}}, \bibinfo
  {author} {\bibfnamefont {T.~R.}\ \bibnamefont {Forrest}}, \bibinfo {author}
  {\bibfnamefont {M.}~\bibnamefont {Abdel-Hafiez}}, \bibinfo {author}
  {\bibfnamefont {X.}~\bibnamefont {Chen}}, \bibinfo {author} {\bibfnamefont
  {D.~A.}\ \bibnamefont {Chareev}}, \bibinfo {author} {\bibfnamefont {A.~N.}\
  \bibnamefont {Vasiliev}}, \bibinfo {author} {\bibfnamefont {P.}~\bibnamefont
  {Bourges}}, \bibinfo {author} {\bibfnamefont {Y.}~\bibnamefont {Sidis}},
  \bibinfo {author} {\bibfnamefont {H.}~\bibnamefont {Cao}}, \ and\ \bibinfo
  {author} {\bibfnamefont {J.}~\bibnamefont {Zhao}},\ }\href {\doibase
  10.1038/nmat4492} {\bibfield  {journal} {\bibinfo  {journal} {Nature
  Materials}\ }\textbf {\bibinfo {volume} {15}},\ \bibinfo {pages} {159}
  (\bibinfo {year} {2016})}\BibitemShut {NoStop}%
\bibitem [{\citenamefont {Kothapalli}\ \emph {et~al.}(2016)\citenamefont
  {Kothapalli}, \citenamefont {B{\"o}hmer}, \citenamefont {Jayasekara},
  \citenamefont {Ueland}, \citenamefont {Das}, \citenamefont {Sapkota},
  \citenamefont {Taufour}, \citenamefont {Xiao}, \citenamefont {Alp},
  \citenamefont {Bud{\~O}ko}, \citenamefont {Canfield}, \citenamefont
  {Kreyssig},\ and\ \citenamefont {Goldman}}]{Kothapalli16}%
  \BibitemOpen
  \bibfield  {author} {\bibinfo {author} {\bibfnamefont {K.}~\bibnamefont
  {Kothapalli}}, \bibinfo {author} {\bibfnamefont {A.~E.}\ \bibnamefont
  {B{\"o}hmer}}, \bibinfo {author} {\bibfnamefont {W.}~\bibnamefont
  {Jayasekara}}, \bibinfo {author} {\bibfnamefont {B.~G.}\ \bibnamefont
  {Ueland}}, \bibinfo {author} {\bibfnamefont {P.}~\bibnamefont {Das}},
  \bibinfo {author} {\bibfnamefont {A.}~\bibnamefont {Sapkota}}, \bibinfo
  {author} {\bibfnamefont {V.}~\bibnamefont {Taufour}}, \bibinfo {author}
  {\bibfnamefont {Y.}~\bibnamefont {Xiao}}, \bibinfo {author} {\bibfnamefont
  {E.}~\bibnamefont {Alp}}, \bibinfo {author} {\bibfnamefont {S.~L.}\
  \bibnamefont {Bud{\~O}ko}}, \bibinfo {author} {\bibfnamefont
  {P.}~\bibnamefont {Canfield}}, \bibinfo {author} {\bibfnamefont
  {A.}~\bibnamefont {Kreyssig}}, \ and\ \bibinfo {author} {\bibfnamefont
  {A.}~\bibnamefont {Goldman}},\ }\href@noop {} {\bibfield  {journal} {\bibinfo
   {journal} {Nature communications}\ }\textbf {\bibinfo {volume} {7}},\
  \bibinfo {pages} {1} (\bibinfo {year} {2016})}\BibitemShut {NoStop}%
\bibitem [{\citenamefont {Matsuura}\ \emph {et~al.}(2017)\citenamefont
  {Matsuura}, \citenamefont {Mizukami}, \citenamefont {Arai}, \citenamefont
  {Sugimura}, \citenamefont {Maejima}, \citenamefont {Machida}, \citenamefont
  {Watanuki}, \citenamefont {Fukuda}, \citenamefont {Yajima}, \citenamefont
  {Hiroi}, \citenamefont {Yip}, \citenamefont {Chan}, \citenamefont {Niu},
  \citenamefont {Hosoi}, \citenamefont {Ishida}, \citenamefont {Mukasa},
  \citenamefont {Kasahara}, \citenamefont {Cheng}, \citenamefont {Goh},
  \citenamefont {Matsuda}, \citenamefont {Uwatoko},\ and\ \citenamefont
  {Shibauchi}}]{Matsuura2017}%
  \BibitemOpen
  \bibfield  {author} {\bibinfo {author} {\bibfnamefont {K.}~\bibnamefont
  {Matsuura}}, \bibinfo {author} {\bibfnamefont {Y.}~\bibnamefont {Mizukami}},
  \bibinfo {author} {\bibfnamefont {Y.}~\bibnamefont {Arai}}, \bibinfo {author}
  {\bibfnamefont {Y.}~\bibnamefont {Sugimura}}, \bibinfo {author}
  {\bibfnamefont {N.}~\bibnamefont {Maejima}}, \bibinfo {author} {\bibfnamefont
  {A.}~\bibnamefont {Machida}}, \bibinfo {author} {\bibfnamefont
  {T.}~\bibnamefont {Watanuki}}, \bibinfo {author} {\bibfnamefont
  {T.}~\bibnamefont {Fukuda}}, \bibinfo {author} {\bibfnamefont
  {T.}~\bibnamefont {Yajima}}, \bibinfo {author} {\bibfnamefont
  {Z.}~\bibnamefont {Hiroi}}, \bibinfo {author} {\bibfnamefont {K.~Y.}\
  \bibnamefont {Yip}}, \bibinfo {author} {\bibfnamefont {Y.~C.}\ \bibnamefont
  {Chan}}, \bibinfo {author} {\bibfnamefont {Q.}~\bibnamefont {Niu}}, \bibinfo
  {author} {\bibfnamefont {S.}~\bibnamefont {Hosoi}}, \bibinfo {author}
  {\bibfnamefont {K.}~\bibnamefont {Ishida}}, \bibinfo {author} {\bibfnamefont
  {K.}~\bibnamefont {Mukasa}}, \bibinfo {author} {\bibfnamefont
  {S.}~\bibnamefont {Kasahara}}, \bibinfo {author} {\bibfnamefont {J.-G.}\
  \bibnamefont {Cheng}}, \bibinfo {author} {\bibfnamefont {S.~K.}\ \bibnamefont
  {Goh}}, \bibinfo {author} {\bibfnamefont {Y.}~\bibnamefont {Matsuda}},
  \bibinfo {author} {\bibfnamefont {Y.}~\bibnamefont {Uwatoko}}, \ and\
  \bibinfo {author} {\bibfnamefont {T.}~\bibnamefont {Shibauchi}},\ }\href
  {\doibase 10.1038/s41467-017-01277-x} {\bibfield  {journal} {\bibinfo
  {journal} {Nature Communications}\ }\textbf {\bibinfo {volume} {8}},\
  \bibinfo {pages} {1143} (\bibinfo {year} {2017})}\BibitemShut {NoStop}%
\bibitem [{\citenamefont {Glasbrenner}\ \emph {et~al.}(2015)\citenamefont
  {Glasbrenner}, \citenamefont {Mazin}, \citenamefont {Jeschke}, \citenamefont
  {Hirschfeld}, \citenamefont {Fernandes},\ and\ \citenamefont
  {Valent{\'i}}}]{Glasbrenner2015}%
  \BibitemOpen
  \bibfield  {author} {\bibinfo {author} {\bibfnamefont {J.~K.}\ \bibnamefont
  {Glasbrenner}}, \bibinfo {author} {\bibfnamefont {I.~I.}\ \bibnamefont
  {Mazin}}, \bibinfo {author} {\bibfnamefont {H.~O.}\ \bibnamefont {Jeschke}},
  \bibinfo {author} {\bibfnamefont {P.~J.}\ \bibnamefont {Hirschfeld}},
  \bibinfo {author} {\bibfnamefont {R.~M.}\ \bibnamefont {Fernandes}}, \ and\
  \bibinfo {author} {\bibfnamefont {R.}~\bibnamefont {Valent{\'i}}},\ }\href
  {\doibase 10.1038/nphys3434} {\bibfield  {journal} {\bibinfo  {journal}
  {Nature Physics}\ }\textbf {\bibinfo {volume} {11}},\ \bibinfo {pages} {953}
  (\bibinfo {year} {2015})}\BibitemShut {NoStop}%
\bibitem [{\citenamefont {Wang}\ \emph {et~al.}(2015)\citenamefont {Wang},
  \citenamefont {Kivelson},\ and\ \citenamefont {Lee}}]{Wang2015}%
  \BibitemOpen
  \bibfield  {author} {\bibinfo {author} {\bibfnamefont {F.}~\bibnamefont
  {Wang}}, \bibinfo {author} {\bibfnamefont {S.~A.}\ \bibnamefont {Kivelson}},
  \ and\ \bibinfo {author} {\bibfnamefont {D.-H.}\ \bibnamefont {Lee}},\ }\href
  {\doibase 10.1038/nphys3456} {\bibfield  {journal} {\bibinfo  {journal}
  {Nature Physics}\ }\textbf {\bibinfo {volume} {11}},\ \bibinfo {pages} {959}
  (\bibinfo {year} {2015})}\BibitemShut {NoStop}%
\bibitem [{\citenamefont {Yu}\ and\ \citenamefont {Si}(2015)}]{Yu2015}%
  \BibitemOpen
  \bibfield  {author} {\bibinfo {author} {\bibfnamefont {R.}~\bibnamefont
  {Yu}}\ and\ \bibinfo {author} {\bibfnamefont {Q.}~\bibnamefont {Si}},\ }\href
  {\doibase 10.1103/PhysRevLett.115.116401} {\bibfield  {journal} {\bibinfo
  {journal} {Phys. Rev. Lett.}\ }\textbf {\bibinfo {volume} {115}},\ \bibinfo
  {pages} {116401} (\bibinfo {year} {2015})}\BibitemShut {NoStop}%
\bibitem [{\citenamefont {Chubukov}\ \emph {et~al.}(2016)\citenamefont
  {Chubukov}, \citenamefont {Khodas},\ and\ \citenamefont
  {Fernandes}}]{Khodas16}%
  \BibitemOpen
  \bibfield  {author} {\bibinfo {author} {\bibfnamefont {A.~V.}\ \bibnamefont
  {Chubukov}}, \bibinfo {author} {\bibfnamefont {M.}~\bibnamefont {Khodas}}, \
  and\ \bibinfo {author} {\bibfnamefont {R.~M.}\ \bibnamefont {Fernandes}},\
  }\href {\doibase 10.1103/PhysRevX.6.041045} {\bibfield  {journal} {\bibinfo
  {journal} {Phys. Rev. X}\ }\textbf {\bibinfo {volume} {6}},\ \bibinfo {pages}
  {041045} (\bibinfo {year} {2016})}\BibitemShut {NoStop}%
\bibitem [{\citenamefont {Yamakawa}\ and\ \citenamefont
  {Kontani}(2017)}]{Kontani17}%
  \BibitemOpen
  \bibfield  {author} {\bibinfo {author} {\bibfnamefont {Y.}~\bibnamefont
  {Yamakawa}}\ and\ \bibinfo {author} {\bibfnamefont {H.}~\bibnamefont
  {Kontani}},\ }\href {\doibase 10.1103/PhysRevB.96.144509} {\bibfield
  {journal} {\bibinfo  {journal} {Phys. Rev. B}\ }\textbf {\bibinfo {volume}
  {96}},\ \bibinfo {pages} {144509} (\bibinfo {year} {2017})}\BibitemShut
  {NoStop}%
\bibitem [{\citenamefont {Terao}\ \emph {et~al.}(2019)\citenamefont {Terao},
  \citenamefont {Kashiwagi}, \citenamefont {Shizu}, \citenamefont {Klemm},\
  and\ \citenamefont {Kadowaki}}]{Terao2019}%
  \BibitemOpen
  \bibfield  {author} {\bibinfo {author} {\bibfnamefont {K.}~\bibnamefont
  {Terao}}, \bibinfo {author} {\bibfnamefont {T.}~\bibnamefont {Kashiwagi}},
  \bibinfo {author} {\bibfnamefont {T.}~\bibnamefont {Shizu}}, \bibinfo
  {author} {\bibfnamefont {R.~A.}\ \bibnamefont {Klemm}}, \ and\ \bibinfo
  {author} {\bibfnamefont {K.}~\bibnamefont {Kadowaki}},\ }\href {\doibase
  10.1103/PhysRevB.100.224516} {\bibfield  {journal} {\bibinfo  {journal}
  {Phys. Rev. B}\ }\textbf {\bibinfo {volume} {100}},\ \bibinfo {pages}
  {224516} (\bibinfo {year} {2019})}\BibitemShut {NoStop}%
\bibitem [{\citenamefont {Lumsden}\ \emph {et~al.}(2010)\citenamefont
  {Lumsden}, \citenamefont {Christianson}, \citenamefont {Goremychkin},
  \citenamefont {Nagler}, \citenamefont {Mook}, \citenamefont {Stone},
  \citenamefont {Abernathy}, \citenamefont {Guidi}, \citenamefont {MacDougall},
  \citenamefont {de~la Cruz}, \citenamefont {Sefat}, \citenamefont {McGuire},
  \citenamefont {Sales},\ and\ \citenamefont {Mandrus}}]{Lumsden2010}%
  \BibitemOpen
  \bibfield  {author} {\bibinfo {author} {\bibfnamefont {M.~D.}\ \bibnamefont
  {Lumsden}}, \bibinfo {author} {\bibfnamefont {A.~D.}\ \bibnamefont
  {Christianson}}, \bibinfo {author} {\bibfnamefont {E.~A.}\ \bibnamefont
  {Goremychkin}}, \bibinfo {author} {\bibfnamefont {S.~E.}\ \bibnamefont
  {Nagler}}, \bibinfo {author} {\bibfnamefont {H.~A.}\ \bibnamefont {Mook}},
  \bibinfo {author} {\bibfnamefont {M.~B.}\ \bibnamefont {Stone}}, \bibinfo
  {author} {\bibfnamefont {D.~L.}\ \bibnamefont {Abernathy}}, \bibinfo {author}
  {\bibfnamefont {T.}~\bibnamefont {Guidi}}, \bibinfo {author} {\bibfnamefont
  {G.~J.}\ \bibnamefont {MacDougall}}, \bibinfo {author} {\bibfnamefont
  {C.}~\bibnamefont {de~la Cruz}}, \bibinfo {author} {\bibfnamefont {A.~S.}\
  \bibnamefont {Sefat}}, \bibinfo {author} {\bibfnamefont {M.~A.}\ \bibnamefont
  {McGuire}}, \bibinfo {author} {\bibfnamefont {B.~C.}\ \bibnamefont {Sales}},
  \ and\ \bibinfo {author} {\bibfnamefont {D.}~\bibnamefont {Mandrus}},\ }\href
  {\doibase 10.1038/nphys1512} {\bibfield  {journal} {\bibinfo  {journal}
  {Nature Physics}\ }\textbf {\bibinfo {volume} {6}},\ \bibinfo {pages} {182}
  (\bibinfo {year} {2010})}\BibitemShut {NoStop}%
\bibitem [{\citenamefont {Liu}\ \emph {et~al.}(2010)\citenamefont {Liu},
  \citenamefont {Hu}, \citenamefont {Qian}, \citenamefont {Fobes},
  \citenamefont {Mao}, \citenamefont {Bao}, \citenamefont {Reehuis},
  \citenamefont {Kimber}, \citenamefont {Proke{\v{s}}}, \citenamefont {Matas},
  \citenamefont {Argyriou}, \citenamefont {Hiess}, \citenamefont {Rotaru},
  \citenamefont {Pham}, \citenamefont {Spinu}, \citenamefont {Qiu},
  \citenamefont {Thampy}, \citenamefont {Savici}, \citenamefont {Rodriguez},\
  and\ \citenamefont {Broholm}}]{Liu2010}%
  \BibitemOpen
  \bibfield  {author} {\bibinfo {author} {\bibfnamefont {T.~J.}\ \bibnamefont
  {Liu}}, \bibinfo {author} {\bibfnamefont {J.}~\bibnamefont {Hu}}, \bibinfo
  {author} {\bibfnamefont {B.}~\bibnamefont {Qian}}, \bibinfo {author}
  {\bibfnamefont {D.}~\bibnamefont {Fobes}}, \bibinfo {author} {\bibfnamefont
  {Z.~Q.}\ \bibnamefont {Mao}}, \bibinfo {author} {\bibfnamefont
  {W.}~\bibnamefont {Bao}}, \bibinfo {author} {\bibfnamefont {M.}~\bibnamefont
  {Reehuis}}, \bibinfo {author} {\bibfnamefont {S.~A.~J.}\ \bibnamefont
  {Kimber}}, \bibinfo {author} {\bibfnamefont {K.}~\bibnamefont
  {Proke{\v{s}}}}, \bibinfo {author} {\bibfnamefont {S.}~\bibnamefont {Matas}},
  \bibinfo {author} {\bibfnamefont {D.~N.}\ \bibnamefont {Argyriou}}, \bibinfo
  {author} {\bibfnamefont {A.}~\bibnamefont {Hiess}}, \bibinfo {author}
  {\bibfnamefont {A.}~\bibnamefont {Rotaru}}, \bibinfo {author} {\bibfnamefont
  {H.}~\bibnamefont {Pham}}, \bibinfo {author} {\bibfnamefont {L.}~\bibnamefont
  {Spinu}}, \bibinfo {author} {\bibfnamefont {Y.}~\bibnamefont {Qiu}}, \bibinfo
  {author} {\bibfnamefont {V.}~\bibnamefont {Thampy}}, \bibinfo {author}
  {\bibfnamefont {A.~T.}\ \bibnamefont {Savici}}, \bibinfo {author}
  {\bibfnamefont {J.~A.}\ \bibnamefont {Rodriguez}}, \ and\ \bibinfo {author}
  {\bibfnamefont {C.}~\bibnamefont {Broholm}},\ }\href {\doibase
  10.1038/nmat2800} {\bibfield  {journal} {\bibinfo  {journal} {Nature
  Materials}\ }\textbf {\bibinfo {volume} {9}},\ \bibinfo {pages} {718}
  (\bibinfo {year} {2010})}\BibitemShut {NoStop}%
\bibitem [{\citenamefont {Zaliznyak}\ \emph {et~al.}(2015)\citenamefont
  {Zaliznyak}, \citenamefont {Savici}, \citenamefont {Lumsden}, \citenamefont
  {Tsvelik}, \citenamefont {Hu},\ and\ \citenamefont
  {Petrovic}}]{Zaliznyak10316}%
  \BibitemOpen
  \bibfield  {author} {\bibinfo {author} {\bibfnamefont {I.}~\bibnamefont
  {Zaliznyak}}, \bibinfo {author} {\bibfnamefont {A.~T.}\ \bibnamefont
  {Savici}}, \bibinfo {author} {\bibfnamefont {M.}~\bibnamefont {Lumsden}},
  \bibinfo {author} {\bibfnamefont {A.}~\bibnamefont {Tsvelik}}, \bibinfo
  {author} {\bibfnamefont {R.}~\bibnamefont {Hu}}, \ and\ \bibinfo {author}
  {\bibfnamefont {C.}~\bibnamefont {Petrovic}},\ }\href {\doibase
  10.1073/pnas.1503559112} {\bibfield  {journal} {\bibinfo  {journal}
  {Proceedings of the National Academy of Sciences}\ }\textbf {\bibinfo
  {volume} {112}},\ \bibinfo {pages} {10316} (\bibinfo {year}
  {2015})}\BibitemShut {NoStop}%
\bibitem [{\citenamefont {Xu}\ \emph {et~al.}(2018)\citenamefont {Xu},
  \citenamefont {Schneeloch}, \citenamefont {Yi}, \citenamefont {Zhao},
  \citenamefont {Matsuda}, \citenamefont {Pajerowski}, \citenamefont {Chi},
  \citenamefont {Birgeneau}, \citenamefont {Gu}, \citenamefont {Tranquada},\
  and\ \citenamefont {Xu}}]{Xu2018}%
  \BibitemOpen
  \bibfield  {author} {\bibinfo {author} {\bibfnamefont {Z.}~\bibnamefont
  {Xu}}, \bibinfo {author} {\bibfnamefont {J.~A.}\ \bibnamefont {Schneeloch}},
  \bibinfo {author} {\bibfnamefont {M.}~\bibnamefont {Yi}}, \bibinfo {author}
  {\bibfnamefont {Y.}~\bibnamefont {Zhao}}, \bibinfo {author} {\bibfnamefont
  {M.}~\bibnamefont {Matsuda}}, \bibinfo {author} {\bibfnamefont {D.~M.}\
  \bibnamefont {Pajerowski}}, \bibinfo {author} {\bibfnamefont
  {S.}~\bibnamefont {Chi}}, \bibinfo {author} {\bibfnamefont {R.~J.}\
  \bibnamefont {Birgeneau}}, \bibinfo {author} {\bibfnamefont {G.}~\bibnamefont
  {Gu}}, \bibinfo {author} {\bibfnamefont {J.~M.}\ \bibnamefont {Tranquada}}, \
  and\ \bibinfo {author} {\bibfnamefont {G.}~\bibnamefont {Xu}},\ }\href
  {\doibase 10.1103/PhysRevB.97.214511} {\bibfield  {journal} {\bibinfo
  {journal} {Phys. Rev. B}\ }\textbf {\bibinfo {volume} {97}},\ \bibinfo
  {pages} {214511} (\bibinfo {year} {2018})}\BibitemShut {NoStop}%
\bibitem [{\citenamefont {Li}\ \emph {et~al.}(2009)\citenamefont {Li},
  \citenamefont {de~la Cruz}, \citenamefont {Huang}, \citenamefont {Chen},
  \citenamefont {Lynn}, \citenamefont {Hu}, \citenamefont {Huang},
  \citenamefont {Hsu}, \citenamefont {Yeh}, \citenamefont {Wu},\ and\
  \citenamefont {Dai}}]{Li2009}%
  \BibitemOpen
  \bibfield  {author} {\bibinfo {author} {\bibfnamefont {S.}~\bibnamefont
  {Li}}, \bibinfo {author} {\bibfnamefont {C.}~\bibnamefont {de~la Cruz}},
  \bibinfo {author} {\bibfnamefont {Q.}~\bibnamefont {Huang}}, \bibinfo
  {author} {\bibfnamefont {Y.}~\bibnamefont {Chen}}, \bibinfo {author}
  {\bibfnamefont {J.~W.}\ \bibnamefont {Lynn}}, \bibinfo {author}
  {\bibfnamefont {J.}~\bibnamefont {Hu}}, \bibinfo {author} {\bibfnamefont
  {Y.-L.}\ \bibnamefont {Huang}}, \bibinfo {author} {\bibfnamefont {F.-C.}\
  \bibnamefont {Hsu}}, \bibinfo {author} {\bibfnamefont {K.-W.}\ \bibnamefont
  {Yeh}}, \bibinfo {author} {\bibfnamefont {M.-K.}\ \bibnamefont {Wu}}, \ and\
  \bibinfo {author} {\bibfnamefont {P.}~\bibnamefont {Dai}},\ }\href {\doibase
  10.1103/PhysRevB.79.054503} {\bibfield  {journal} {\bibinfo  {journal} {Phys.
  Rev. B}\ }\textbf {\bibinfo {volume} {79}},\ \bibinfo {pages} {054503}
  (\bibinfo {year} {2009})}\BibitemShut {NoStop}%
\bibitem [{\citenamefont {Kuo}\ \emph {et~al.}(2016)\citenamefont {Kuo},
  \citenamefont {Chu}, \citenamefont {Palmstrom}, \citenamefont {Kivelson},\
  and\ \citenamefont {Fisher}}]{Kuo958}%
  \BibitemOpen
  \bibfield  {author} {\bibinfo {author} {\bibfnamefont {H.-H.}\ \bibnamefont
  {Kuo}}, \bibinfo {author} {\bibfnamefont {J.-H.}\ \bibnamefont {Chu}},
  \bibinfo {author} {\bibfnamefont {J.~C.}\ \bibnamefont {Palmstrom}}, \bibinfo
  {author} {\bibfnamefont {S.~A.}\ \bibnamefont {Kivelson}}, \ and\ \bibinfo
  {author} {\bibfnamefont {I.~R.}\ \bibnamefont {Fisher}},\ }\href {\doibase
  10.1126/science.aab0103} {\bibfield  {journal} {\bibinfo  {journal}
  {Science}\ }\textbf {\bibinfo {volume} {352}},\ \bibinfo {pages} {958}
  (\bibinfo {year} {2016})}\BibitemShut {NoStop}%
\bibitem [{\citenamefont {Yin}\ \emph {et~al.}(2011)\citenamefont {Yin},
  \citenamefont {Haule},\ and\ \citenamefont {Kotliar}}]{Yin2011}%
  \BibitemOpen
  \bibfield  {author} {\bibinfo {author} {\bibfnamefont {Z.~P.}\ \bibnamefont
  {Yin}}, \bibinfo {author} {\bibfnamefont {K.}~\bibnamefont {Haule}}, \ and\
  \bibinfo {author} {\bibfnamefont {G.}~\bibnamefont {Kotliar}},\ }\href
  {\doibase 10.1038/nmat3120} {\bibfield  {journal} {\bibinfo  {journal}
  {Nature Materials}\ }\textbf {\bibinfo {volume} {10}},\ \bibinfo {pages}
  {932} (\bibinfo {year} {2011})}\BibitemShut {NoStop}%
\bibitem [{\citenamefont {Lanat\`a}\ \emph {et~al.}(2013)\citenamefont
  {Lanat\`a}, \citenamefont {Strand}, \citenamefont {Giovannetti},
  \citenamefont {Hellsing}, \citenamefont {de' Medici},\ and\ \citenamefont
  {Capone}}]{Lanata2013}%
  \BibitemOpen
  \bibfield  {author} {\bibinfo {author} {\bibfnamefont {N.}~\bibnamefont
  {Lanat\`a}}, \bibinfo {author} {\bibfnamefont {H.~U.~R.}\ \bibnamefont
  {Strand}}, \bibinfo {author} {\bibfnamefont {G.}~\bibnamefont {Giovannetti}},
  \bibinfo {author} {\bibfnamefont {B.}~\bibnamefont {Hellsing}}, \bibinfo
  {author} {\bibfnamefont {L.}~\bibnamefont {de' Medici}}, \ and\ \bibinfo
  {author} {\bibfnamefont {M.}~\bibnamefont {Capone}},\ }\href {\doibase
  10.1103/PhysRevB.87.045122} {\bibfield  {journal} {\bibinfo  {journal} {Phys.
  Rev. B}\ }\textbf {\bibinfo {volume} {87}},\ \bibinfo {pages} {045122}
  (\bibinfo {year} {2013})}\BibitemShut {NoStop}%
\bibitem [{\citenamefont {de' Medici}\ \emph {et~al.}(2014)\citenamefont {de'
  Medici}, \citenamefont {Giovannetti},\ and\ \citenamefont
  {Capone}}]{Medici2014}%
  \BibitemOpen
  \bibfield  {author} {\bibinfo {author} {\bibfnamefont {L.}~\bibnamefont {de'
  Medici}}, \bibinfo {author} {\bibfnamefont {G.}~\bibnamefont {Giovannetti}},
  \ and\ \bibinfo {author} {\bibfnamefont {M.}~\bibnamefont {Capone}},\ }\href
  {\doibase 10.1103/PhysRevLett.112.177001} {\bibfield  {journal} {\bibinfo
  {journal} {Phys. Rev. Lett.}\ }\textbf {\bibinfo {volume} {112}},\ \bibinfo
  {pages} {177001} (\bibinfo {year} {2014})}\BibitemShut {NoStop}%
\bibitem [{\citenamefont {Sprau}\ \emph {et~al.}(2017)\citenamefont {Sprau},
  \citenamefont {Kostin}, \citenamefont {Kreisel}, \citenamefont {B{\"o}hmer},
  \citenamefont {Taufour}, \citenamefont {Canfield}, \citenamefont {Mukherjee},
  \citenamefont {Hirschfeld}, \citenamefont {Andersen},\ and\ \citenamefont
  {Davis}}]{Sprau75}%
  \BibitemOpen
  \bibfield  {author} {\bibinfo {author} {\bibfnamefont {P.~O.}\ \bibnamefont
  {Sprau}}, \bibinfo {author} {\bibfnamefont {A.}~\bibnamefont {Kostin}},
  \bibinfo {author} {\bibfnamefont {A.}~\bibnamefont {Kreisel}}, \bibinfo
  {author} {\bibfnamefont {A.~E.}\ \bibnamefont {B{\"o}hmer}}, \bibinfo
  {author} {\bibfnamefont {V.}~\bibnamefont {Taufour}}, \bibinfo {author}
  {\bibfnamefont {P.~C.}\ \bibnamefont {Canfield}}, \bibinfo {author}
  {\bibfnamefont {S.}~\bibnamefont {Mukherjee}}, \bibinfo {author}
  {\bibfnamefont {P.~J.}\ \bibnamefont {Hirschfeld}}, \bibinfo {author}
  {\bibfnamefont {B.~M.}\ \bibnamefont {Andersen}}, \ and\ \bibinfo {author}
  {\bibfnamefont {J.~C.~S.}\ \bibnamefont {Davis}},\ }\href {\doibase
  10.1126/science.aal1575} {\bibfield  {journal} {\bibinfo  {journal}
  {Science}\ }\textbf {\bibinfo {volume} {357}},\ \bibinfo {pages} {75}
  (\bibinfo {year} {2017})}\BibitemShut {NoStop}%
\bibitem [{\citenamefont {Fanfarillo}\ \emph {et~al.}(2018)\citenamefont
  {Fanfarillo}, \citenamefont {Benfatto},\ and\ \citenamefont
  {Valenzuela}}]{Fanfarillo2018}%
  \BibitemOpen
  \bibfield  {author} {\bibinfo {author} {\bibfnamefont {L.}~\bibnamefont
  {Fanfarillo}}, \bibinfo {author} {\bibfnamefont {L.}~\bibnamefont
  {Benfatto}}, \ and\ \bibinfo {author} {\bibfnamefont {B.}~\bibnamefont
  {Valenzuela}},\ }\href {\doibase 10.1103/PhysRevB.97.121109} {\bibfield
  {journal} {\bibinfo  {journal} {Phys. Rev. B}\ }\textbf {\bibinfo {volume}
  {97}},\ \bibinfo {pages} {121109} (\bibinfo {year} {2018})}\BibitemShut
  {NoStop}%
\bibitem [{\citenamefont {Yu}\ \emph {et~al.}(2018)\citenamefont {Yu},
  \citenamefont {Zhu},\ and\ \citenamefont {Si}}]{Yu2018}%
  \BibitemOpen
  \bibfield  {author} {\bibinfo {author} {\bibfnamefont {R.}~\bibnamefont
  {Yu}}, \bibinfo {author} {\bibfnamefont {J.-X.}\ \bibnamefont {Zhu}}, \ and\
  \bibinfo {author} {\bibfnamefont {Q.}~\bibnamefont {Si}},\ }\href {\doibase
  10.1103/PhysRevLett.121.227003} {\bibfield  {journal} {\bibinfo  {journal}
  {Phys. Rev. Lett.}\ }\textbf {\bibinfo {volume} {121}},\ \bibinfo {pages}
  {227003} (\bibinfo {year} {2018})}\BibitemShut {NoStop}%
\bibitem [{\citenamefont {Yi}\ \emph {et~al.}(2015)\citenamefont {Yi},
  \citenamefont {Liu}, \citenamefont {Zhang}, \citenamefont {Yu}, \citenamefont
  {Zhu}, \citenamefont {Lee}, \citenamefont {Moore}, \citenamefont {Schmitt},
  \citenamefont {Li}, \citenamefont {Riggs}, \citenamefont {Chu}, \citenamefont
  {Lv}, \citenamefont {Hu}, \citenamefont {Hashimoto}, \citenamefont {Mo},
  \citenamefont {Hussain}, \citenamefont {Mao}, \citenamefont {Chu},
  \citenamefont {Fisher}, \citenamefont {Si}, \citenamefont {Shen},\ and\
  \citenamefont {Lu}}]{Yi2015}%
  \BibitemOpen
  \bibfield  {author} {\bibinfo {author} {\bibfnamefont {M.}~\bibnamefont
  {Yi}}, \bibinfo {author} {\bibfnamefont {Z.-K.}\ \bibnamefont {Liu}},
  \bibinfo {author} {\bibfnamefont {Y.}~\bibnamefont {Zhang}}, \bibinfo
  {author} {\bibfnamefont {R.}~\bibnamefont {Yu}}, \bibinfo {author}
  {\bibfnamefont {J.-X.}\ \bibnamefont {Zhu}}, \bibinfo {author} {\bibfnamefont
  {J.~J.}\ \bibnamefont {Lee}}, \bibinfo {author} {\bibfnamefont {R.~G.}\
  \bibnamefont {Moore}}, \bibinfo {author} {\bibfnamefont {F.~T.}\ \bibnamefont
  {Schmitt}}, \bibinfo {author} {\bibfnamefont {W.}~\bibnamefont {Li}},
  \bibinfo {author} {\bibfnamefont {S.~C.}\ \bibnamefont {Riggs}}, \bibinfo
  {author} {\bibfnamefont {J.-H.}\ \bibnamefont {Chu}}, \bibinfo {author}
  {\bibfnamefont {B.}~\bibnamefont {Lv}}, \bibinfo {author} {\bibfnamefont
  {J.}~\bibnamefont {Hu}}, \bibinfo {author} {\bibfnamefont {M.}~\bibnamefont
  {Hashimoto}}, \bibinfo {author} {\bibfnamefont {S.-K.}\ \bibnamefont {Mo}},
  \bibinfo {author} {\bibfnamefont {Z.}~\bibnamefont {Hussain}}, \bibinfo
  {author} {\bibfnamefont {Z.~Q.}\ \bibnamefont {Mao}}, \bibinfo {author}
  {\bibfnamefont {C.~W.}\ \bibnamefont {Chu}}, \bibinfo {author} {\bibfnamefont
  {I.~R.}\ \bibnamefont {Fisher}}, \bibinfo {author} {\bibfnamefont
  {Q.}~\bibnamefont {Si}}, \bibinfo {author} {\bibfnamefont {Z.-X.}\
  \bibnamefont {Shen}}, \ and\ \bibinfo {author} {\bibfnamefont {D.~H.}\
  \bibnamefont {Lu}},\ }\href {\doibase 10.1038/ncomms8777} {\bibfield
  {journal} {\bibinfo  {journal} {Nature Communications}\ }\textbf {\bibinfo
  {volume} {6}},\ \bibinfo {pages} {7777} (\bibinfo {year} {2015})}\BibitemShut
  {NoStop}%
\bibitem [{\citenamefont {Liu}\ \emph {et~al.}(2015)\citenamefont {Liu},
  \citenamefont {Yi}, \citenamefont {Zhang}, \citenamefont {Hu}, \citenamefont
  {Yu}, \citenamefont {Zhu}, \citenamefont {He}, \citenamefont {Chen},
  \citenamefont {Hashimoto}, \citenamefont {Moore}, \citenamefont {Mo},
  \citenamefont {Hussain}, \citenamefont {Si}, \citenamefont {Mao},
  \citenamefont {Lu},\ and\ \citenamefont {Shen}}]{Liu2015}%
  \BibitemOpen
  \bibfield  {author} {\bibinfo {author} {\bibfnamefont {Z.~K.}\ \bibnamefont
  {Liu}}, \bibinfo {author} {\bibfnamefont {M.}~\bibnamefont {Yi}}, \bibinfo
  {author} {\bibfnamefont {Y.}~\bibnamefont {Zhang}}, \bibinfo {author}
  {\bibfnamefont {J.}~\bibnamefont {Hu}}, \bibinfo {author} {\bibfnamefont
  {R.}~\bibnamefont {Yu}}, \bibinfo {author} {\bibfnamefont {J.-X.}\
  \bibnamefont {Zhu}}, \bibinfo {author} {\bibfnamefont {R.-H.}\ \bibnamefont
  {He}}, \bibinfo {author} {\bibfnamefont {Y.~L.}\ \bibnamefont {Chen}},
  \bibinfo {author} {\bibfnamefont {M.}~\bibnamefont {Hashimoto}}, \bibinfo
  {author} {\bibfnamefont {R.~G.}\ \bibnamefont {Moore}}, \bibinfo {author}
  {\bibfnamefont {S.-K.}\ \bibnamefont {Mo}}, \bibinfo {author} {\bibfnamefont
  {Z.}~\bibnamefont {Hussain}}, \bibinfo {author} {\bibfnamefont
  {Q.}~\bibnamefont {Si}}, \bibinfo {author} {\bibfnamefont {Z.~Q.}\
  \bibnamefont {Mao}}, \bibinfo {author} {\bibfnamefont {D.~H.}\ \bibnamefont
  {Lu}}, \ and\ \bibinfo {author} {\bibfnamefont {Z.-X.}\ \bibnamefont
  {Shen}},\ }\href {\doibase 10.1103/PhysRevB.92.235138} {\bibfield  {journal}
  {\bibinfo  {journal} {Phys. Rev. B}\ }\textbf {\bibinfo {volume} {92}},\
  \bibinfo {pages} {235138} (\bibinfo {year} {2015})}\BibitemShut {NoStop}%
\bibitem [{\citenamefont {Fanfarillo}\ \emph {et~al.}(2017)\citenamefont
  {Fanfarillo}, \citenamefont {Giovannetti}, \citenamefont {Capone},\ and\
  \citenamefont {Bascones}}]{Bascones17}%
  \BibitemOpen
  \bibfield  {author} {\bibinfo {author} {\bibfnamefont {L.}~\bibnamefont
  {Fanfarillo}}, \bibinfo {author} {\bibfnamefont {G.}~\bibnamefont
  {Giovannetti}}, \bibinfo {author} {\bibfnamefont {M.}~\bibnamefont {Capone}},
  \ and\ \bibinfo {author} {\bibfnamefont {E.}~\bibnamefont {Bascones}},\
  }\href {\doibase 10.1103/PhysRevB.95.144511} {\bibfield  {journal} {\bibinfo
  {journal} {Phys. Rev. B}\ }\textbf {\bibinfo {volume} {95}},\ \bibinfo
  {pages} {144511} (\bibinfo {year} {2017})}\BibitemShut {NoStop}%
\bibitem [{\citenamefont {Zhang}\ \emph {et~al.}(2017)\citenamefont {Zhang},
  \citenamefont {Glasbrenner}, \citenamefont {Flint}, \citenamefont {Mazin},\
  and\ \citenamefont {Fernandes}}]{Zhang2017}%
  \BibitemOpen
  \bibfield  {author} {\bibinfo {author} {\bibfnamefont {G.}~\bibnamefont
  {Zhang}}, \bibinfo {author} {\bibfnamefont {J.~K.}\ \bibnamefont
  {Glasbrenner}}, \bibinfo {author} {\bibfnamefont {R.}~\bibnamefont {Flint}},
  \bibinfo {author} {\bibfnamefont {I.~I.}\ \bibnamefont {Mazin}}, \ and\
  \bibinfo {author} {\bibfnamefont {R.~M.}\ \bibnamefont {Fernandes}},\ }\href
  {\doibase 10.1103/PhysRevB.95.174402} {\bibfield  {journal} {\bibinfo
  {journal} {Phys. Rev. B}\ }\textbf {\bibinfo {volume} {95}},\ \bibinfo
  {pages} {174402} (\bibinfo {year} {2017})}\BibitemShut {NoStop}%
\bibitem [{\citenamefont {Bishop}\ \emph {et~al.}(2017)\citenamefont {Bishop},
  \citenamefont {Herbrych}, \citenamefont {Dagotto},\ and\ \citenamefont
  {Moreo}}]{Moreo17}%
  \BibitemOpen
  \bibfield  {author} {\bibinfo {author} {\bibfnamefont {C.~B.}\ \bibnamefont
  {Bishop}}, \bibinfo {author} {\bibfnamefont {J.}~\bibnamefont {Herbrych}},
  \bibinfo {author} {\bibfnamefont {E.}~\bibnamefont {Dagotto}}, \ and\
  \bibinfo {author} {\bibfnamefont {A.}~\bibnamefont {Moreo}},\ }\href
  {\doibase 10.1103/PhysRevB.96.035144} {\bibfield  {journal} {\bibinfo
  {journal} {Phys. Rev. B}\ }\textbf {\bibinfo {volume} {96}},\ \bibinfo
  {pages} {035144} (\bibinfo {year} {2017})}\BibitemShut {NoStop}%
\bibitem [{\citenamefont {Borisov}\ \emph {et~al.}(2019)\citenamefont
  {Borisov}, \citenamefont {Fernandes},\ and\ \citenamefont
  {Valent\'{\i}}}]{Borisov2019}%
  \BibitemOpen
  \bibfield  {author} {\bibinfo {author} {\bibfnamefont {V.}~\bibnamefont
  {Borisov}}, \bibinfo {author} {\bibfnamefont {R.~M.}\ \bibnamefont
  {Fernandes}}, \ and\ \bibinfo {author} {\bibfnamefont {R.}~\bibnamefont
  {Valent\'{\i}}},\ }\href {\doibase 10.1103/PhysRevLett.123.146402} {\bibfield
   {journal} {\bibinfo  {journal} {Phys. Rev. Lett.}\ }\textbf {\bibinfo
  {volume} {123}},\ \bibinfo {pages} {146402} (\bibinfo {year}
  {2019})}\BibitemShut {NoStop}%
\bibitem [{\citenamefont {Ding}\ \emph {et~al.}(2014)\citenamefont {Ding},
  \citenamefont {Pan}, \citenamefont {Yang},\ and\ \citenamefont
  {Wen}}]{Ding2014}%
  \BibitemOpen
  \bibfield  {author} {\bibinfo {author} {\bibfnamefont {X.}~\bibnamefont
  {Ding}}, \bibinfo {author} {\bibfnamefont {Y.}~\bibnamefont {Pan}}, \bibinfo
  {author} {\bibfnamefont {H.}~\bibnamefont {Yang}}, \ and\ \bibinfo {author}
  {\bibfnamefont {H.-H.}\ \bibnamefont {Wen}},\ }\href {\doibase
  10.1103/PhysRevB.89.224515} {\bibfield  {journal} {\bibinfo  {journal} {Phys.
  Rev. B}\ }\textbf {\bibinfo {volume} {89}},\ \bibinfo {pages} {224515}
  (\bibinfo {year} {2014})}\BibitemShut {NoStop}%
\bibitem [{\citenamefont {Otsuka}\ \emph {et~al.}(2019)\citenamefont {Otsuka},
  \citenamefont {Hagisawa}, \citenamefont {Koshika}, \citenamefont {Adachi},
  \citenamefont {Usui}, \citenamefont {Sasaki}, \citenamefont {Sasaki},
  \citenamefont {Yamaguchi}, \citenamefont {Nakanishi}, \citenamefont
  {Yoshizawa}, \citenamefont {Kimura},\ and\ \citenamefont
  {Watanabe}}]{Otsuka2019}%
  \BibitemOpen
  \bibfield  {author} {\bibinfo {author} {\bibfnamefont {T.}~\bibnamefont
  {Otsuka}}, \bibinfo {author} {\bibfnamefont {S.}~\bibnamefont {Hagisawa}},
  \bibinfo {author} {\bibfnamefont {Y.}~\bibnamefont {Koshika}}, \bibinfo
  {author} {\bibfnamefont {S.}~\bibnamefont {Adachi}}, \bibinfo {author}
  {\bibfnamefont {T.}~\bibnamefont {Usui}}, \bibinfo {author} {\bibfnamefont
  {N.}~\bibnamefont {Sasaki}}, \bibinfo {author} {\bibfnamefont
  {S.}~\bibnamefont {Sasaki}}, \bibinfo {author} {\bibfnamefont
  {S.}~\bibnamefont {Yamaguchi}}, \bibinfo {author} {\bibfnamefont
  {Y.}~\bibnamefont {Nakanishi}}, \bibinfo {author} {\bibfnamefont
  {M.}~\bibnamefont {Yoshizawa}}, \bibinfo {author} {\bibfnamefont
  {S.}~\bibnamefont {Kimura}}, \ and\ \bibinfo {author} {\bibfnamefont
  {T.}~\bibnamefont {Watanabe}},\ }\href {\doibase 10.1103/PhysRevB.99.184505}
  {\bibfield  {journal} {\bibinfo  {journal} {Phys. Rev. B}\ }\textbf {\bibinfo
  {volume} {99}},\ \bibinfo {pages} {184505} (\bibinfo {year}
  {2019})}\BibitemShut {NoStop}%
\bibitem [{\citenamefont {Sun}\ \emph {et~al.}(2019)\citenamefont {Sun},
  \citenamefont {Shi},\ and\ \citenamefont {Tamegai}}]{Sun_2019}%
  \BibitemOpen
  \bibfield  {author} {\bibinfo {author} {\bibfnamefont {Y.}~\bibnamefont
  {Sun}}, \bibinfo {author} {\bibfnamefont {Z.}~\bibnamefont {Shi}}, \ and\
  \bibinfo {author} {\bibfnamefont {T.}~\bibnamefont {Tamegai}},\ }\href
  {\doibase 10.1088/1361-6668/ab30c2} {\bibfield  {journal} {\bibinfo
  {journal} {Superconductor Science and Technology}\ }\textbf {\bibinfo
  {volume} {32}},\ \bibinfo {pages} {103001} (\bibinfo {year}
  {2019})}\BibitemShut {NoStop}%
\bibitem [{\citenamefont {Fanfarillo}\ \emph {et~al.}(2015)\citenamefont
  {Fanfarillo}, \citenamefont {Cortijo},\ and\ \citenamefont
  {Valenzuela}}]{Fanfarillo15}%
  \BibitemOpen
  \bibfield  {author} {\bibinfo {author} {\bibfnamefont {L.}~\bibnamefont
  {Fanfarillo}}, \bibinfo {author} {\bibfnamefont {A.}~\bibnamefont {Cortijo}},
  \ and\ \bibinfo {author} {\bibfnamefont {B.}~\bibnamefont {Valenzuela}},\
  }\href {\doibase 10.1103/PhysRevB.91.214515} {\bibfield  {journal} {\bibinfo
  {journal} {Phys. Rev. B}\ }\textbf {\bibinfo {volume} {91}},\ \bibinfo
  {pages} {214515} (\bibinfo {year} {2015})}\BibitemShut {NoStop}%
\bibitem [{\citenamefont {Onari}\ \emph {et~al.}(2016)\citenamefont {Onari},
  \citenamefont {Yamakawa},\ and\ \citenamefont {Kontani}}]{Kontani16}%
  \BibitemOpen
  \bibfield  {author} {\bibinfo {author} {\bibfnamefont {S.}~\bibnamefont
  {Onari}}, \bibinfo {author} {\bibfnamefont {Y.}~\bibnamefont {Yamakawa}}, \
  and\ \bibinfo {author} {\bibfnamefont {H.}~\bibnamefont {Kontani}},\ }\href
  {\doibase 10.1103/PhysRevLett.116.227001} {\bibfield  {journal} {\bibinfo
  {journal} {Phys. Rev. Lett.}\ }\textbf {\bibinfo {volume} {116}},\ \bibinfo
  {pages} {227001} (\bibinfo {year} {2016})}\BibitemShut {NoStop}%
\bibitem [{\citenamefont {Xing}\ \emph {et~al.}(2018)\citenamefont {Xing},
  \citenamefont {Classen},\ and\ \citenamefont {Chubukov}}]{Chubukov18}%
  \BibitemOpen
  \bibfield  {author} {\bibinfo {author} {\bibfnamefont {R.-Q.}\ \bibnamefont
  {Xing}}, \bibinfo {author} {\bibfnamefont {L.}~\bibnamefont {Classen}}, \
  and\ \bibinfo {author} {\bibfnamefont {A.~V.}\ \bibnamefont {Chubukov}},\
  }\href {\doibase 10.1103/PhysRevB.98.041108} {\bibfield  {journal} {\bibinfo
  {journal} {Phys. Rev. B}\ }\textbf {\bibinfo {volume} {98}},\ \bibinfo
  {pages} {041108} (\bibinfo {year} {2018})}\BibitemShut {NoStop}%
\bibitem [{\citenamefont {Christensen}\ \emph {et~al.}(2016)\citenamefont
  {Christensen}, \citenamefont {Kang}, \citenamefont {Andersen},\ and\
  \citenamefont {Fernandes}}]{Christensen2016}%
  \BibitemOpen
  \bibfield  {author} {\bibinfo {author} {\bibfnamefont {M.~H.}\ \bibnamefont
  {Christensen}}, \bibinfo {author} {\bibfnamefont {J.}~\bibnamefont {Kang}},
  \bibinfo {author} {\bibfnamefont {B.~M.}\ \bibnamefont {Andersen}}, \ and\
  \bibinfo {author} {\bibfnamefont {R.~M.}\ \bibnamefont {Fernandes}},\ }\href
  {\doibase 10.1103/PhysRevB.93.085136} {\bibfield  {journal} {\bibinfo
  {journal} {Phys. Rev. B}\ }\textbf {\bibinfo {volume} {93}},\ \bibinfo
  {pages} {085136} (\bibinfo {year} {2016})}\BibitemShut {NoStop}%
\bibitem [{\citenamefont {Xu}\ \emph {et~al.}(2012)\citenamefont {Xu},
  \citenamefont {Wen}, \citenamefont {Zhao}, \citenamefont {Matsuda},
  \citenamefont {Ku}, \citenamefont {Liu}, \citenamefont {Gu}, \citenamefont
  {Lee}, \citenamefont {Birgeneau}, \citenamefont {Tranquada},\ and\
  \citenamefont {Xu}}]{Xu2012}%
  \BibitemOpen
  \bibfield  {author} {\bibinfo {author} {\bibfnamefont {Z.}~\bibnamefont
  {Xu}}, \bibinfo {author} {\bibfnamefont {J.}~\bibnamefont {Wen}}, \bibinfo
  {author} {\bibfnamefont {Y.}~\bibnamefont {Zhao}}, \bibinfo {author}
  {\bibfnamefont {M.}~\bibnamefont {Matsuda}}, \bibinfo {author} {\bibfnamefont
  {W.}~\bibnamefont {Ku}}, \bibinfo {author} {\bibfnamefont {X.}~\bibnamefont
  {Liu}}, \bibinfo {author} {\bibfnamefont {G.}~\bibnamefont {Gu}}, \bibinfo
  {author} {\bibfnamefont {D.-H.}\ \bibnamefont {Lee}}, \bibinfo {author}
  {\bibfnamefont {R.~J.}\ \bibnamefont {Birgeneau}}, \bibinfo {author}
  {\bibfnamefont {J.~M.}\ \bibnamefont {Tranquada}}, \ and\ \bibinfo {author}
  {\bibfnamefont {G.}~\bibnamefont {Xu}},\ }\href {\doibase
  10.1103/PhysRevLett.109.227002} {\bibfield  {journal} {\bibinfo  {journal}
  {Phys. Rev. Lett.}\ }\textbf {\bibinfo {volume} {109}},\ \bibinfo {pages}
  {227002} (\bibinfo {year} {2012})}\BibitemShut {NoStop}%
\end{thebibliography}%
\end{document}

\title{Supplementary Material for Nematic Fluctuations in an Orbital Selective Superconductor Fe$_{1+y}$Te$_{1-x}$Se$_{x}$}
	\author{Qianni Jiang}
	\affiliation{Department of Physics, University of Washington, Seattle, Washington 98195, USA}
	\author{Yue Shi}
	\affiliation{Department of Physics, University of Washington, Seattle, Washington 98195, USA}
	\author{Morten Christensen}
	\affiliation{School of Physics and Astronomy, University of Minnesota, Minneapolis, Minnesota 55455, USA}
	\author{Joshua Sanchez}
	\affiliation{Department of Physics, University of Washington, Seattle, Washington 98195, USA}
	\author{Bevin Huang}
	\affiliation{Department of Physics, University of Washington, Seattle, Washington 98195, USA}
	\author{Zhong Lin}
	\affiliation{Department of Physics, University of Washington, Seattle, Washington 98195, USA}
	\author{Zhaoyu Liu}
	\affiliation{Department of Physics, University of Washington, Seattle, Washington 98195, USA}
	\author{Paul Malinowski}
	\affiliation{Department of Physics, University of Washington, Seattle, Washington 98195, USA}
	\author{Xiaodong Xu}
	\affiliation{Department of Physics, University of Washington, Seattle, Washington 98195, USA}
	\affiliation{Department of Material Science and Engineering, University of Washington, Seattle, Washington 98195, USA}
	\author{Rafael Fernandes}
	\affiliation{School of Physics and Astronomy, University of Minnesota, Minneapolis, Minnesota 55455, USA}
	\author{Jiun-Haw Chu}
	\affiliation{Department of Physics, University of Washington, Seattle, Washington 98195, USA}
	\date{\today}
	\pacs{}
	\maketitle
	\titlespacing\section{0pt}{18pt plus 4pt minus 2pt}{18pt plus 2pt minus 2pt}
	\section{S1 Polarized Raman spectroscopy on Fe$_{1+y}$Te$_{0.55}$Se$_{0.45}$}
\begin{figure*}
	\renewcommand{\figurename}{Fig. S}
	\includegraphics[trim={0 0.1cm 0cm 0.2cm},clip,width=1\textwidth]{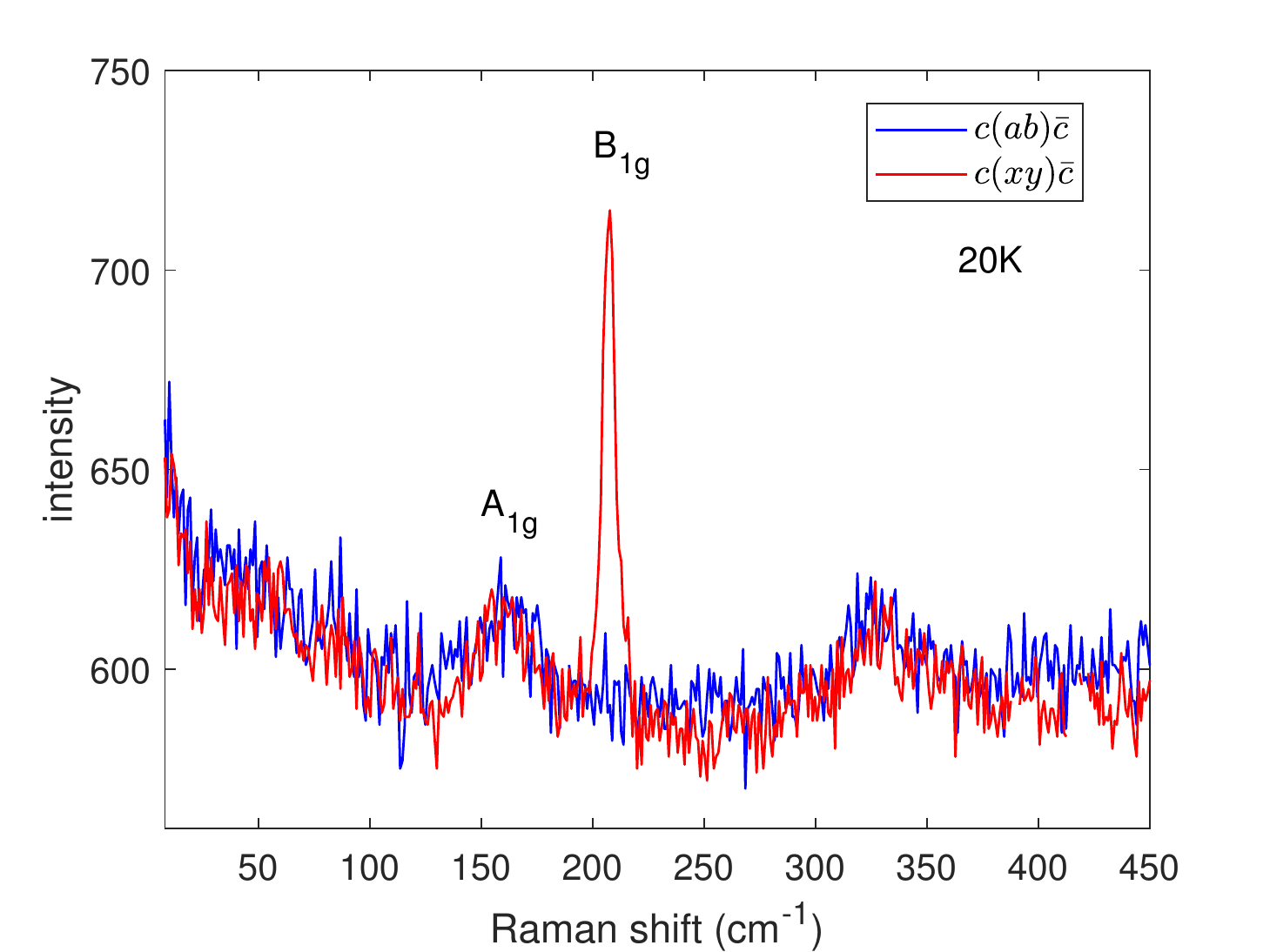}
	\caption{Polarized Raman spectra on  Fe$_{1+y}$Te$_{0.55}$Se$_{0.45}$ in the (ab) and (xy) configurations at 20K. The $B_{1g}$ mode at 208 cm$^{-1}$ distinguishes the (xy) configuration from the (ab) configuration.}
	\label{fig:FigS1}
\end{figure*}

	Polarized Raman spectroscopy was used to determine the crystal orientation. Representative data of the Raman spectra are shown in FIG. S\ref{fig:FigS1}. Measurements were made on freshly cleaved surfaces of annealed Fe$_{1+y}$Te$_{0.55}$Se$_{0.45}$ single crystals in a backscattering configuration at 20K using a 532nm laser, a 1200 groove mm$^{-1}$ grating spectrometer (Princeton Acton 2500i) and a liquid-nitrogen cooled charge-coupled device (CCD) detector. The laser power was set to 100$\mu$W and the laser spot was focused on a $\sim$ 2 $\times$ 2 $\mu m^{2}$ surface. The (ab) configuration corresponds to the polarization of the incident (scattered) light along the crystallographic a (b) axis, while the (xy) configuration corresponds to the polarization of the incident (scattered) light along the x (y) direction, which is the direction rotated by 45$^{\circ}$ from the a (b) axis in the ab plane. The presence and absence of $B_{1g}$ phonon mode at 208 cm$^{-1}$ in the (xy) and (ab)  configuration unambiguously confirm the determination of crystal orientation. A weak $A_{1g}$  phonon mode at 159 cm$^{-1}$ was observed in both (ab) and (xy) configurations due to a leakage of other polarization components as observed elsewhere. 
\section{S2 Annealing effects of Fe$_{1+y}$Te$_{1-x}$Se$_{x}$}
\begin{figure*}
	\includegraphics[trim={0 0cm 0 0.2cm},clip,width=1\textwidth]{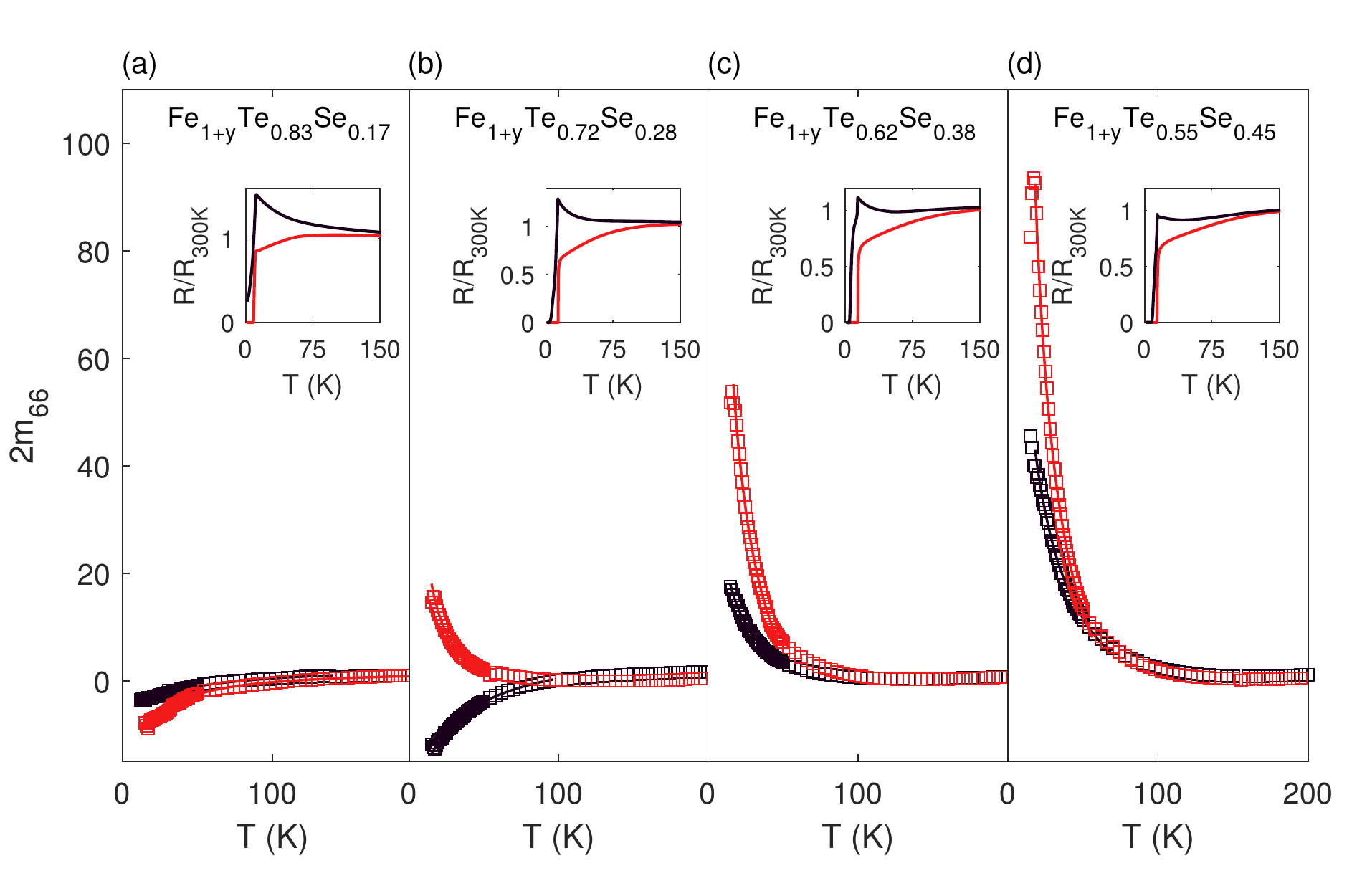}
	\renewcommand{\figurename}{Fig. S}
	\caption{Temperature dependence of $2m_{66}$ for as-grown (black squares) and annealed (red squares) (a) Fe$_{1+y}$Te$_{0.83}$Se$_{0.17}$ (b) Fe$_{1+y}$Te$_{0.72}$Se$_{0.28}$ (c) Fe$_{1+y}$Te$_{0.62}$Se$_{0.38}$ and (d) Fe$_{1+y}$Te$_{0.55}$Se$_{0.45}$. The black and red lines underneath the square data show the Curie Weiss fittings for the $2m_{66}$ of as-grown and annealed samples respectively. The fitting parameters are listed in Table S\ref{tab:tabs2}. For as-grown crystals, $2m_{66}$ for x = 0.17 and 0.28 are negative. After annealing, $2m_{66}$ for x = 0.28 flips sign from negative to positive and the $2m_{66}$ for x = 0.17 remains negative. Insets of (a)-(d) are the temperature dependences of normalized resistances for as-grown (black) and annealed (red) Fe$_{1+y}$Te$_{1-x}$Se$_{x}$ (x = 0.17 0.28 0.38 0.45) respectively.}
	\label{fig:figS3}
\end{figure*}

As grown crystals are known to host excess iron which can be removed by annealing in Te or Se vapor. Samples with and without excess Fe show drastically different electrical, magnetic, and superconducting properties. To study these effects, crystals were cleaved into thin slices ($\sim$ 1 mm), loaded in a crucible with another crucible of an appropriate amount of selenium powder beneath it, sealed in quartz tubes, and annealed at 500$^\circ$C for a week. Further annealing in a Se vapor causes significant hardship in cleaving sizable samples. The average compositions of Fe$_{1+y}$Te$_{1-x}$Se$_{x}$ samples were determined on at least 4 regions of the crystals by the energy-dispersive x-ray (EDXS) method using the Sirion XL30 scanning electron microscope. We only use EDS measurement to determine x (Se/Te) ratio. For the measurement of y, even though we consistently obtained a smaller value of y after annealing, the determination of the absolute value of y is inconclusive. Similar difficulty to accurately determined y has also been reported previously \cite{Rinott17}. By comparing the transport data and the position and the width of transition, we estimate the upper limit of y is below 0.06 for x $<$ 0.4 and and 0.02 for x $>$ 0.4. 
\\
\indent The annealing significantly changed the electrical transport behavior, as shown in FIG. S\ref{fig:figS3} insets. After annealing, the resistivity changes from a weakly insulating behavior for as-grown crystals (black curves) to a metallic behaivor for annealed samples (red curves). The superconducting transition also becomes sharper with an increased T$_{c}$ after annealing. Fig. S\ref{fig:figS3} shows the elastoresistivity coefficient in the B$_{2g}$ channel, $2m_{66}$, for both as-grown and annealed Fe$_{1+y}$Te$_{1-x}$Se$_{x}$ (x = 0.17 - 0.45). The elastoresistivity coefficient $2m_{66}$ is enhanced for the annealed samples (red squares), suggesting the correlation between the enhanced nematic susceptibility and the coherence of the $d_{xy}$ orbital as discussed in the main text. We notice that the $2m_{66}$ changes sign as x decreases, and the x = 0.28 even shows a sign changing before and after annealing. This behavior might be related to the sensitivity of the sign of resistivity anisotropy to the shape of Fermi surfaces and spin fluctuations, which warrants future investigation.
\section{S3 Elastoresistivity measurement}
\begin{figure}
	\includegraphics[trim={0 0cm 0cm 0cm},clip,width=0.45\textwidth]{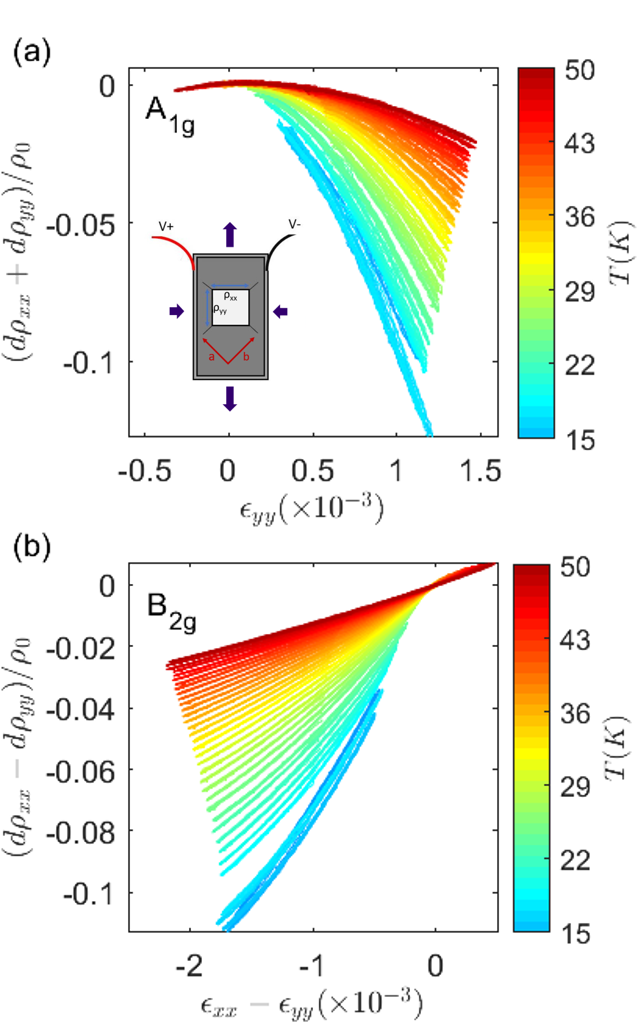}
	\renewcommand{\figurename}{Fig. S}
	\caption{Temperature dependence of (a) the isotropic ($A_{1g}$) elastoresistivity response to the tensile strain and (b) the anisotropic ($B_{2g}$) elastoresistivity response to the anisotropic strain, of a single crystal of Fe$_{1+y}$Te$_{0.55}$Se$_{0.45}$ prepared along the Fe-Fe bonding direction. The zero-strain point is determined by the intersection of the strain dependence of $\rho_{xx}$ and $\rho_{yy}$ at each temperature. }
	\label{fig:FigS4}
\end{figure}
\indent The elastoresistivity measurement is conducted by gluing the square shape samples on a piezoelectric stack with the Montgomery contact geometry. Square edge of the samples are cut along the Fe-Fe (Fe-Ch) bonding direction for $B_{2g}$ ($B_{1g}$) elastoresistivity measurements, with electrical contacts at the four corners of the square. With this method, we are able to measure the resistivity along two perpendicular directions on the same piece of sample, and therefore eliminate the problems caused by variations of the samples and the transmitted strain. By gluing the sample on the sidewall of the piezoelectric stack and applying voltages, we apply an in situ tunable anisotropic strain on the sample. The magnitude of the strain is calibrated by measuring the resistance change of a strain gauge glued on the other side of the piezo stack. In this approach, one can decompose the shear strain into different symmetry channels and precisely determine the elastoresistivity coefficients. In the configuration shown in the inset of FIG. S\ref{fig:FigS4}(a), the strain can be decomposed into the anisotropic $B_{2g}$ strain and the isotropic $A_{1g}$ strain. FIG. S\ref{fig:FigS4}(a) and (b) show the elastoresistivity responses in the $A_{1g}$ and $B_{2g}$ symmetry channels. The linear slope at zero strain point of the elastoresistivity in the $A_{1g}$ and $B_{2g}$ channels are the first order elastoresistivity coefficients $m_{A_{1g}}^{A_{1g}}$ and $m_{B_{2g}}^{B_{2g}}$, respectively. One may also notice nonlinear responses in the elastoresistivity of annealed Fe$_{1+y}$Te$_{0.55}$Se$_{0.45}$ are present in both the $B_{2g}$ and $A_{1g}$ channels. According to the symmetry constraints, the nonlinear term in the $A_{1g}$ channel is most likely due to the second-order $B_{2g}$ strain:
\begin{eqnarray}
(\frac{\Delta\rho}{\rho_{0}})_{A_{1g}}=m_{A_{1g}}^{A_{1g}}\epsilon_{A_{1g}}+m_{A_{1g}}^{B_{2g},B_{2g}}(\epsilon_{B_{2g}})^{2}+m_{A_{1g}}^{A_{1g},A_{1g}}(\epsilon_{A_{1g}})^{2}+O(\epsilon^{3})
\end{eqnarray}
\indent While for the $B_{2g}$ response (FIG. S\ref{fig:FigS4} (b)), the nonlinearity in the $B_{2g}$ channel can be caused by either a mixed-in isotropic resistivity due to the sample's deviation from a perfect square along the Fe-Fe bonding direction, or a second-order strain $\epsilon_{B_{2g}}\epsilon_{A_{1g}}$. 
\begin{eqnarray}
(\frac{\Delta\rho}{\rho_{0}})_{B_{2g}}^{m}=m_{B_{2g}}^{B_{2g}}\epsilon_{B_{2g}}+m_{A_{1g}}^{B_{2g},B_{2g},mixed}(\epsilon_{B_{2g}})^{2}+m_{B_{2g}}^{B_{2g},A_{1g}}\epsilon_{B_{2g}}\epsilon_{A_{1g}}+O(\epsilon^{3})
\end{eqnarray}	
\indent Here, $(\frac{\Delta\rho}{\rho_{0}})_{B_{2g}}^{m}$ is the measured value of resistivity anisotropy, which contains the mixed-in isotropic component. In this measurement, we mainly focus on the anisotropy term $m_{B_{2g}}^{B_{2g}}$, also known as $2m_{66}$ in the Voigt notation, which is proportional to the nematic susceptibility in the $B_{2g}$ symmetry channel of $D_{4h}$ symmetry group. By finding the anisotropic strain neutral point and fitting the anisotropic elastoresistivity response quadratically, we are able to remove the mixed-in error and extract the first order elastoresistivity coefficient $2m_{66}$. 
\section{S4 Curie-Weiss fitting of elastoresistivity coefficient $2m_{66}$ and $m_{11} - m_{12}$}
\begin{figure*}
	\includegraphics[trim={0cm 2cm 0cm 1cm},clip,width=1\textwidth]{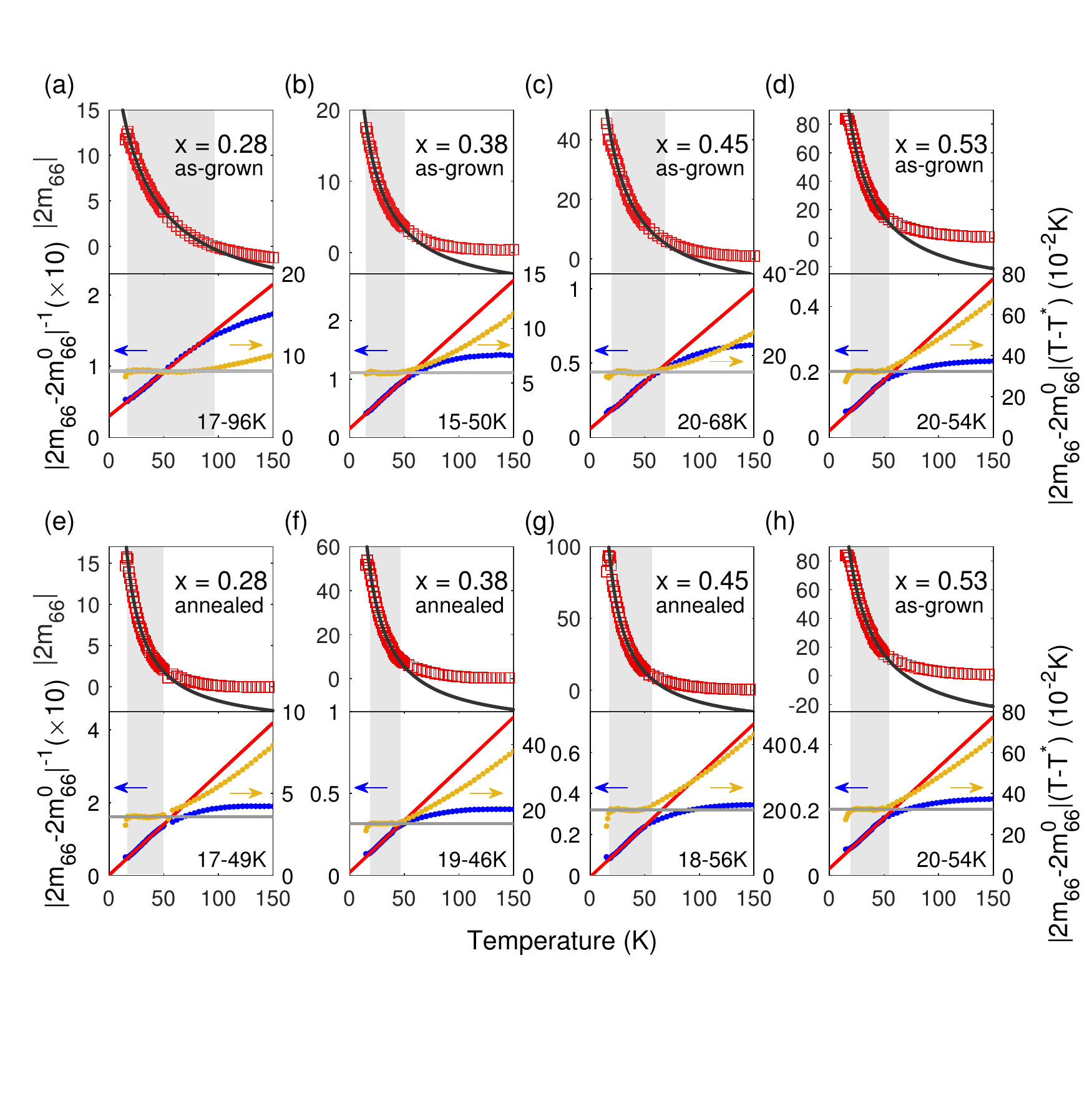}
	\renewcommand{\figurename}{Fig. S}
	\caption{Curie Weiss fitting of the $B_{2g}$ elastoresistivity coefficient $2m_{66}$ of both as-grown and annealed Fe$_{1+y}$Te$_{1-x}$Se$_{x}$ (x =  0.28, 0.38, 0.45, 0.53). (a) - (d) as-grown Fe$_{1+y}$Te$_{1-x}$Se$_{x}$. (e) - (g) annealed Fe$_{1+y}$Te$_{1-x}$Se$_{x}$. Upper panels show $|2m_{66}|$, whereas lower panels show $|2m_{66} - 2m_{66}^{0}|^{-1}$ (left axes of lower panels, blue symbols) and  $|2m_{66} - 2m_{66}^{0}|(T-T^{*})$ (right axes of lower panels, yellow circles). Black (upper panels) and red (lower panels) lines show the fits to Curie-Weiss behavior of $|2m_{66}|$ and  $|2m_{66} - 2m_{66}^{0}|^{-1}$respectively. Grey horizontal lines (lower panels) shows the average values of $|2m_{66} - 2m_{66}^{0}|(T-T^{*})$ in the fitting temperature range. The optimal fitting range (grey shaded) is determined by the greatest corresponding adjusted R-Square value. For (a),  $2m_{66}$ is negative. For (b) - (h),  $2m_{66}$ is positive.}
	\label{fig:FIGS5}
\end{figure*}
\indent FIG. S\ref{fig:FIGS5} shows the Curie Weiss fitting of the $B_{2g}$ elastoresistivity coefficient $2m_{66}$ for both as-grown and annealed Fe$_{1+y}$Te$_{1-x}$Se$_{x}$ (x = 0.28 - 0.53). The optimal fitting range is illustrated by the grey shaded area. The divergent temperature dependence of $2m_{66}$ can be fitted by the Curie Weiss law well below 50K but deviates from the Curie Weiss law at higher temperatures. Other fitting parameters are listed in Table S\ref{tab:tabs2}. The bare nematic critical temperature $T^{*}$ is negative for all as-grown samples. With the Se concentration increases, the bare nematic critical temperature $T^{*}$ increases and approaches zero for x = 0.53. For annealed samples, the $T^{*}$ extracted from the Curie-Weiss fitting is approximately zero, while the Curie constant $\lambda/a$ decreases with x. Near x = 0, in the proximity of the double spin stripe order, a diverging $B_{1g}$ nematic susceptibility was also observed. The Curie Weiss fitting parameters from the fit of $B_{1g}$ elastoresistivity coefficient $m_{11} - m_{12}$ are listed in Table S\ref{tab:tabs3}. 
\squeezetable
\begin{table}
		\renewcommand{\tablename}{TABLE S}
	\caption{\label{tab:tabs2}Curie-Weiss fitting parameters from the Fit of $2m_{66}$ for as-grown and annealed Fe$_{1+y}$Te$_{1-x}$Se$_{x}$ (x = 0.17 - 0.53)}
	\begin{ruledtabular}
		\begin{tabular}{ccccccc}
			Materials&  Fitting Range (K) & $2m_{66}^{0}$ & $\lambda/a_{0} (K)$ & $T^{*} (K)$ &  Sample Dimension ($\mu m$) & Adjust R-square\\
			\hline
			&&&As-grown&&&\\
			\hline
			Fe$_{1+y}$Te$_{0.83}$Se$_{0.17}$ & $11-140$ & $3.2\pm0.4$ & $-397\pm16$ & $-44.6\pm7.7$ & $530\times480\times20$ & $0.9904$\\
			Fe$_{1+y}$Te$_{0.72}$Se$_{0.28}$ & $17-96$ & $7.0\pm0.6$ & $-811\pm10$ & $-24.5\pm2.9$ & $1210\times1150\times10$ & $0.9982$\\
			Fe$_{1+y}$Te$_{0.62}$Se$_{0.38}$ & $15-50$ & $-6.8\pm1.0$ & $592\pm55$ & $-9.0\pm2.1$ & $870\times740\times40$ & $0.9984$\\
			Fe$_{1+y}$Te$_{0.55}$Se$_{0.45}$ & $20-68$ & $-15.22\pm3.0$ & $1594\pm222$ & $-9.3\pm1.2$ & $1080\times1070\times30$ & $0.9966$\\
			Fe$_{1+y}$Te$_{0.47}$Se$_{0.53}$ & $20-93$ & $-16.7\pm2.0$ & $1752\pm159$ & $-0.8\pm1.7$ & $670\times570\times60$ & $0.9981$\\
			\hline
			&&&Annealed&&&\\
			\hline
			Fe$_{1+y}$Te$_{0.83}$Se$_{0.17}$ & $25-200$ & $2.2\pm0.2$ & $-243\pm10$ & $-1.9\pm1.9$ & $460\times440\times10$ & $0.9965$\\
			Fe$_{1+y}$Te$_{0.72}$Se$_{0.28}$ & $17-49$ & $-5.3\pm0.6$ & $359\pm32$ & $-0.4\pm1.2$ & $840\times800\times50$ & $0.9991$\\
			Fe$_{1+y}$Te$_{0.62}$Se$_{0.38}$ & $19-46$ & $-24.4\pm2.3$ & $1583\pm135$ & $-2.9\pm1.5$ & $380\times360\times30$ & $0.9995$\\
			Fe$_{1+y}$Te$_{0.55}$Se$_{0.45}$ & $18-56$ & $-28.6\pm3.1$ & $2004\pm186$ & $1.5\pm1.2$ & $410\times360\times30$ & $0.9988$
		\end{tabular}
	\end{ruledtabular}
\end{table}
\squeezetable
\begin{table}
	\renewcommand{\tablename}{TABLE S}
	\caption{\label{tab:tabs3}Curie-Weiss fitting parameters from the Fit of $m_{11} - m_{12}$ for Fe$_{1+y}$Te}
	\begin{ruledtabular}
		\begin{tabular}{ccccccc}
			Materials&  Fitting Range (K) & $m_{11}^{0}-m_{12}^{0}$ & $\lambda/a_{0} (K)$ & $T^{*} (K)$ &  Sample Dimension ($\mu m$) & Adjust R-square\\
			\hline
			Fe$_{1+y}$Te & $71-104$ & $0.29\pm0.06$ & $13\pm4$ & $70.2\pm0.1$ & $443\times440\times24$ & $0.9992$\\
		\end{tabular}
	\end{ruledtabular}
\end{table}

\section{S5 RPA calculation of the nematic susceptibility}

To capture the effect of orbital degrees of freedom on the nematic susceptibility, we begin by introducing the multi-orbital Hubbard-Kanamori Hamiltonian (momentum sums are implicit):
\begin{eqnarray}
	\mathcal{H} &=& \sum_{\mu} \left[\epsilon_{\mu\nu}(\mbf{k}) - \mu \delta_{\mu\nu}\right]c^{\dagger}_{\mbf{k}\mu\sigma}c_{\mbf{k}\nu\sigma} \nonumber + U \sum_{\mu} n_{\mbf{q}\mu\uparrow} n_{-\mbf{q}\mu\downarrow} + U' \sum_{\substack{\mu < \nu \\ \sigma \sigma'}}n_{\mbf{q}\mu\sigma}n_{-\mbf{q}\nu\sigma'} \\ &+& \frac{J}{2}\sum_{\substack{\mu\neq\nu \\ \sigma\sigma'}}c^{\dagger}_{\mbf{k}+\mbf{q}\mu\sigma}c_{\mbf{k}\nu\sigma}c^{\dagger}_{\mbf{k}'-\mbf{q}\nu\sigma'}c_{\mbf{k}'\mu\sigma} + \frac{J'}{2}\sum_{\substack{\mu\neq\nu \\ \sigma}}c^{\dagger}_{\mbf{k}+\mbf{q}\mu\sigma}c^{\dagger}_{\mbf{k}'-\mbf{q}\mu\bar{\sigma}}c_{\mbf{k}'\nu\bar{\sigma}}c_{\mbf{k}\nu\sigma}\,.
\end{eqnarray}
Here $\mu$ and $\nu$ are orbital indices, $\sigma$ labels spin and we assume $U'=U-2J$ and $J'=J$. $\epsilon^{\mu\nu}(\mbf{k})$ denotes the dispersion, in this case obtained from a tight-binding fit to DFT. In the results shown in the main text, we used the band structure parameters presented in Ref.~[\onlinecite{ikeda10}], which give three hole-like Fermi pockets and two electron-like Fermi pockets. While this tight-binding parametrization is not intended to model specifically optimally doped FeTe$_{1-x}$Se$_x$, it offers a solid framework to elucidate, on general grounds, the tendencies of how the nematic susceptibility is affected by the suppression of $d_{xy}$ orbital spectral weight.

The interactions can be conveniently expressed as elements of the same tensor:
\begin{eqnarray}
	U^{\mu\mu\mu\mu} = U\,, \quad U^{\mu\nu\nu\mu} = U'\,, \quad U^{\mu\nu\mu\nu} = J'\,, \quad U^{\mu\mu\nu\nu} = J\,.
\end{eqnarray}
The RPA expression for the spin-driven nematic susceptibility of this model was derived previously by two of us in Ref. \cite{christensen16} (see also Ref. \cite{fanfarillo18}). Here, we simply quote the result from the former paper, derived in the approximation that the antiferromagnetic order parameter is diagonal in orbital space (as shown for instance in Ref. \cite{christensen17}). The nematic susceptibility in this case is a rank-4 tensor given by:
\begin{eqnarray}
	\chi_{\rm nem}^{\mu\nu\rho\lambda}/N = \left( \int_q \chi^{\rho \iota}(q)\chi^{\lambda \kappa}(q) \right) \left(\delta_{\mu \kappa}\delta_{\nu \iota} - g^{\mu\nu \gamma \phi} \int_q \chi^{\gamma \iota }(q)\chi^{\phi \kappa}(q) \right)^{-1}\,,\label{eq:nem_susc}
\end{eqnarray}
where $\int_q \equiv T \sum_{\omega_m}\int \mathrm{d}^2 \mbf{q}$, $N=3$ is the number of components of the magnetic order parameter, and the Einstein summation convention is used. Here
\begin{eqnarray}
	\chi^{\mu\nu}(q) = \left[ (U_{\mu\nu})^{-1} - \chi_0^{\mu\nu}(q) \right]^{-1}
\end{eqnarray}
is the RPA magnetic susceptibility, which is a rank-2 tensor. The bare magnetic susceptibility is given by:
\begin{eqnarray}
	\chi^{\mu\nu}_{0}(q) = - \sum_k \mathcal{G}^{\mu\nu}(k+q)\mathcal{G}^{\nu\mu}(k)\,,
\end{eqnarray}
where repeated indices are not summed and $\mathcal{G}^{\mu\nu}$ denotes the bare multi-orbital Green's function:
\begin{eqnarray}
	\mathcal{G}^{\mu\nu}(k) = \sum_{m} \frac{a^{m}_{\mu}(\mbf{k})a^{m}_{\nu}(\mbf{k})^{\ast}}{i\omega_n - \xi^{m}(\mbf{k})}\,,
\end{eqnarray}
with $k=(i\omega_m,\mbf{k})$. For notational convenience, we introduce:
\begin{eqnarray}
	\mathcal{G}_X^{\mu\nu} \equiv \mathcal{G}^{\mu\nu}(\mbf{k}+\mbf{Q}_1,i\omega_m)\,,\quad 
	\mathcal{G}_Y^{\mu\nu} \equiv \mathcal{G}^{\mu\nu}(\mbf{k}+\mbf{Q}_2,i\omega_m)\,.
\end{eqnarray}
Here, $\mbf{Q}_1 = (\pi, 0)$ and $\mbf{Q}_2 = (0, \pi)$ (in the 1-Fe unit cell). Finally, the nematic coupling constant $g$ in Eq. (\ref{eq:nem_susc}) is also a rank-4 tensor obtained from convolutions of the Green's functions:
\begin{eqnarray}
	g^{\rho\nu\eta\mu} &=& -\frac{1}{16}\sum_{k}\Big( 2\mathcal{G}^{\mu\rho}\mathcal{G}^{\rho\nu}_{X} \mathcal{G}^{\nu\eta}\mathcal{G}^{\eta\mu}_{X} - \mathcal{G}^{\mu\rho}\mathcal{G}^{\rho\eta}_{X} \mathcal{G}^{\eta\nu}\mathcal{G}^{\nu\mu}_{X} - \mathcal{G}^{\mu\rho}\mathcal{G}^{\rho\nu}_{X} \mathcal{G}^{\nu\eta}\mathcal{G}^{\eta\mu}_{Y} \nonumber \\ && \qquad - \mathcal{G}^{\nu\rho}\mathcal{G}^{\rho\mu}_{X} \mathcal{G}^{\mu\eta}_{X+Y}\mathcal{G}^{\eta\nu}_{X} + \mathcal{G}^{\mu\rho}\mathcal{G}^{\rho\eta}_{X} \mathcal{G}^{\eta\nu}_{X+Y}\mathcal{G}^{\nu\mu}_{Y} \Big) + (X \leftrightarrow Y)\,,
\end{eqnarray}
where repeated indices are not summed.

To elucidate how orbital differentiation affects the nematic susceptibility within our framework, we follow the approach outlined in Ref.~\cite{kreisel17} and modify the Green's function by:
\begin{eqnarray}
	\mathcal{G}^{\mu \nu} \rightarrow \sqrt{Z_{\mu}Z_{\nu}}\mathcal{G}^{\mu\nu}\,,
\end{eqnarray}
where we stress that, once again, the Einstein convention is not assumed. Since both $g^{\mu\nu\rho\lambda}$ and $\chi^{\mu\nu}(q)$ depend on the Green's functions, a reduced $Z$ factor will have an intricate impact on the nematic susceptibility. As discussed in the main text, this is a phenomenological way to mimic the complicated effect of incoherence and loss of spectral weight on the response function, whose validity requires that the system remains in a metallic state.

\begin{figure}
\includegraphics[width=0.6\textwidth]{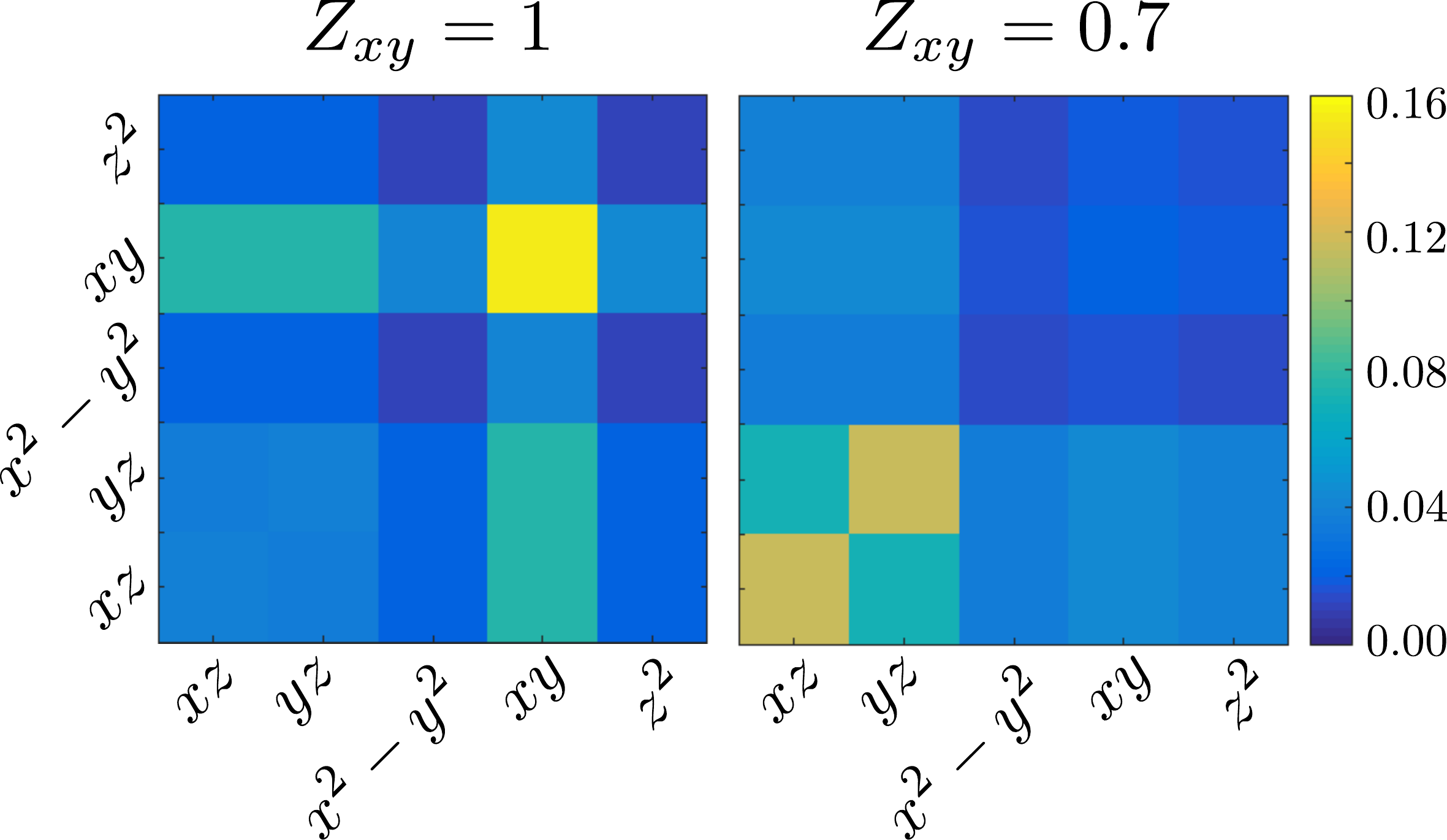}
\renewcommand{\figurename}{Fig. S}
\caption{\label{fig:nematic_susc_weight} $\Phi^{(1)}_{\mu\nu}$ for $Z_{xy}=1$ and $Z_{xy}=0.7$ at a temperature immediately prior to the nematic instability of each case. For $Z_{xy}=1$, the dominant contribution to the nematic susceptibility arises from the $xy$ orbital. As $Z_{xy}$ is reduced, the $xz$ and $yz$ orbitals become the dominant ones.}
\end{figure}

The nematic susceptibility plotted in the main text for different values of $Z_{xy}$ corresponds to the largest eigenvalue $\lambda_{\rm nem}$ of the equation:
\begin{equation}
\chi_{\rm nem}^{\mu\nu\rho\lambda} \Phi^{(n)}_{\mu \nu} = \lambda^{(n)}_{\rm nem} \Phi^{(n)}_{\rho \lambda} 
\end{equation}
In Fig. S\ref{fig:nematic_susc_weight}, we show how the corresponding eigenvector $\Phi^{(n)}_{\mu \nu}$ changes for decreasing $Z_{xy}$. Here we used $U=1.2$ eV and $J=U/6$. To highlight the impact of $Z_{xy}$ we fixed the filling to $5.9$ electrons per site. As expected, the main contribution to the nematic susceptibility, signaled by the brightest squares in the figure, shifts from intra $xy$-orbital processes to intra $xz/yz$-orbital processes. 

\begin{figure}
\includegraphics[width=0.4\textwidth]{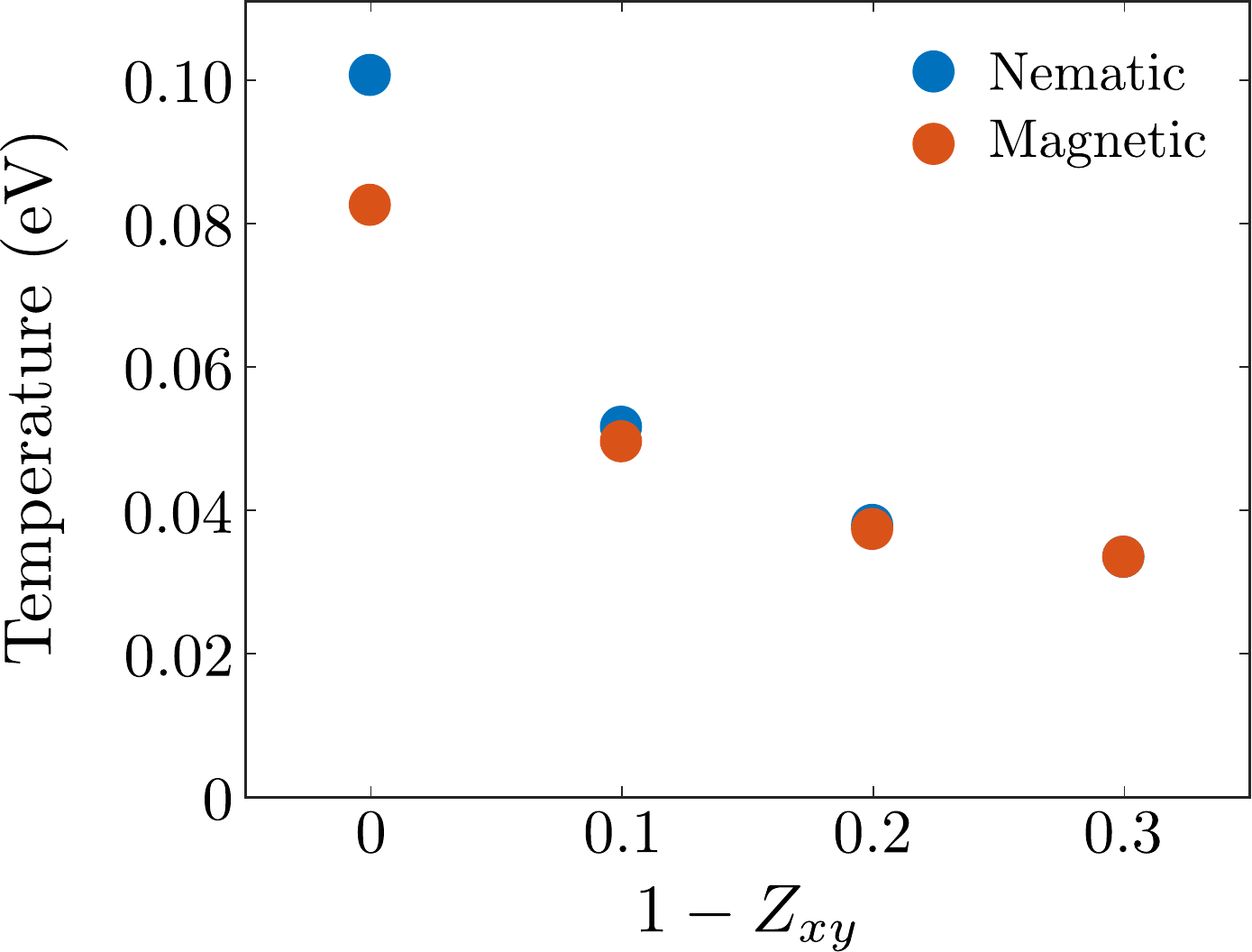}
\renewcommand{\figurename}{Fig. S}
\caption{\label{fig:temp_as_Z} Nematic (blue) and bare magnetic (red) transition temperatures as a function of $1-Z_{xy}$. Both are reduced by reducing $Z_{xy}$, along with their relative separation.  At $Z_{xy}=0.7$ the separation between the two vanishes within our temperature resolution ($\delta T < 0.2$ meV).}
\end{figure}

A direct consequence of the reduction of the nematic susceptibility is a suppression of the nematic transition temperature, which in our model is manifested as a divergence of the largest eigenvalue $\lambda_{\rm nem}$. In Fig. S\ref{fig:temp_as_Z}, we plot both the nematic transition temperature $T_{\rm nem}$ and the bare (i.e. non-renormalized by nematic order) magnetic transition temperature $T_{\rm mag}$ as a function of $Z_{xy}$. Note that not only are both transition temperatures strongly suppressed, but their separation also decreases significantly for decreasing $Z_{xy}$. As noted in Ref. \cite{christensen16}, these RPA transition temperatures are, not surprisingly, overestimated with respect to the actual transition temperatures. For this reason, and to be able to compare the temperature dependencies of the nematic susceptibilities of systems with very different values of $T_{\rm nem}$, in the main text we plot $\lambda_{\rm nem}$ as a function of $T - T_{\rm nem}$.

\begin{figure}
\includegraphics[width=0.5\textwidth]{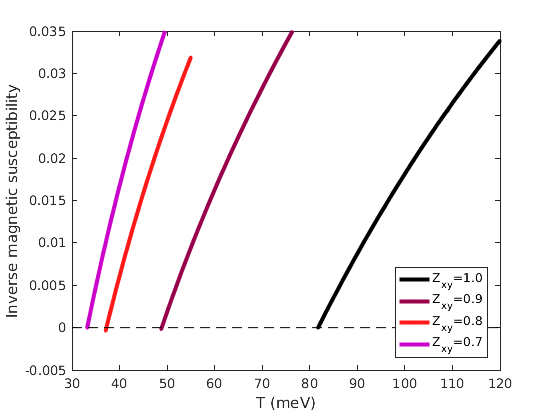}
\renewcommand{\figurename}{Fig. S}
\caption{\label{fig:magnetic_susceptibility} Inverse magnetic susceptibility for $ 0.7 \leq Z_{xy}\leq 1$ as a function of temperature. $Z_{xy}$ has a pronounced effect on the intensity of the magnetic susceptibility for a fixed temperature, and reducing $Z_{xy}$ reduces the overall intensity. On the other hand, the temperature dependence of the susceptibility is little affected by the change in $Z_{xy}$, and the magnetic susceptibility remains Curie-Weiss like, albeit the slope changes. Here we show the leading eigenvalue $\lambda_{\rm mag}$ as defined in Eq.~\eqref{eq:mag_susc_def}.}
\end{figure}

The suppression of $T_{\rm mag}$ for decreasing $Z_{xy}$ is also manifested in the suppression of the overall magnitude of the bare magnetic susceptibility calculated at the spin-stripe wave-vector, $\chi_{\rm mag}$, as shown in Fig. S\ref{fig:magnetic_susceptibility}.
Here, $\chi_{\rm mag}$ was calculated in a manner similar to the nematic susceptibility:
\begin{equation}
	\chi^{\mu\nu}(\mathbf{q}) \Psi^{(n)}_{\nu}(\mathbf{q}) = \lambda_{\rm mag}^{(n)}(\mathbf{q})\Psi^{(n)}_{\mu}(\mathbf{q})\,,\label{eq:mag_susc_def}
\end{equation}
where, in Fig. S\ref{fig:magnetic_susceptibility}, $\mathbf{q}=\mathbf{Q}_1$ or $\mathbf{q}=\mathbf{Q}_2$ and we show the leading eigenvalue. Note that, in contrast to the nematic susceptibility, the temperature dependence of $\chi_{\rm mag}$ is not strongly affected by the reduction of $d_{xy}$ spectral weight, as seen clearly from Fig. S\ref{fig:magnetic_susceptibility}. However, for a fixed temperature, there is a strong suppression of $\chi_{\rm mag}$ with decreasing $Z_{xy}$. This last feature is in qualitative agreement with neutron scattering experiments, which showed that, in optimally-doped FeTe$_{1-x}$Se$_x$, the low-energy stripe-type magnetic fluctuations are suppressed with increasing temperature \cite{xu12}. According to our analysis in the main text, increasing the temperature promotes a less coherent $d_{xy}$ orbital. Note, however, that Ref. \cite{xu12} also reported that, as temperature was increased, besides a suppression of intensity at the stripe wave-vector, an incommensurate peak appeared in the momentum-resolved magnetic susceptibility at low energies. We did not observe such incommensurate peaks in our energy integrated magnetic susceptibility, i.e. the peak remains at $\mathbf{q}=\mathbf{Q}_1$ (and at $\mathbf{q}=\mathbf{Q}_2$), suggesting that this effect cannot be captured phenomenologically by a constant $Z_{xy}$ factor.